\RequirePackage{lineno}
\documentclass[aps,prl,twocolumn,showpacs,superscriptaddress,groupedaddress]{revtex4}  
\usepackage{graphicx}  
\usepackage{dcolumn}   
\usepackage{bm}        
\usepackage{amssymb}   
\usepackage{lineno}

\def \PeTe {$p_T$}
\def \met  {\mbox{$\not\!\!E_T$}}

\begin{document}

\hspace{5.2in} \mbox{Fermilab-Pub-11/313-E}

\title{Search for associated Higgs boson production using like charge dilepton events
       \\ in $\boldsymbol{p\bar{p}}$ collisions at $\boldsymbol{\sqrt{s} =}$ 1.96~TeV}
%
\affiliation{Universidad de Buenos Aires, Buenos Aires, Argentina}
\affiliation{LAFEX, Centro Brasileiro de Pesquisas F{\'\i}sicas, Rio de Janeiro, Brazil}
\affiliation{Universidade do Estado do Rio de Janeiro, Rio de Janeiro, Brazil}
\affiliation{Universidade Federal do ABC, Santo Andr\'e, Brazil}
\affiliation{Instituto de F\'{\i}sica Te\'orica, Universidade Estadual Paulista, S\~ao Paulo, Brazil}
\affiliation{Simon Fraser University, Vancouver, British Columbia, and York University, Toronto, Ontario, Canada}
\affiliation{University of Science and Technology of China, Hefei, People's Republic of China}
\affiliation{Universidad de los Andes, Bogot\'{a}, Colombia}
\affiliation{Charles University, Faculty of Mathematics and Physics, Center for Particle Physics, Prague, Czech Republic}
\affiliation{Czech Technical University in Prague, Prague, Czech Republic}
\affiliation{Center for Particle Physics, Institute of Physics, Academy of Sciences of the Czech Republic, Prague, Czech Republic}
\affiliation{Universidad San Francisco de Quito, Quito, Ecuador}
\affiliation{LPC, Universit\'e Blaise Pascal, CNRS/IN2P3, Clermont, France}
\affiliation{LPSC, Universit\'e Joseph Fourier Grenoble 1, CNRS/IN2P3, Institut National Polytechnique de Grenoble, Grenoble, France}
\affiliation{CPPM, Aix-Marseille Universit\'e, CNRS/IN2P3, Marseille, France}
\affiliation{LAL, Universit\'e Paris-Sud, CNRS/IN2P3, Orsay, France}
\affiliation{LPNHE, Universit\'es Paris VI and VII, CNRS/IN2P3, Paris, France}
\affiliation{CEA, Irfu, SPP, Saclay, France}
\affiliation{IPHC, Universit\'e de Strasbourg, CNRS/IN2P3, Strasbourg, France}
\affiliation{IPNL, Universit\'e Lyon 1, CNRS/IN2P3, Villeurbanne, France and Universit\'e de Lyon, Lyon, France}
\affiliation{III. Physikalisches Institut A, RWTH Aachen University, Aachen, Germany}
\affiliation{Physikalisches Institut, Universit{\"a}t Freiburg, Freiburg, Germany}
\affiliation{II. Physikalisches Institut, Georg-August-Universit{\"a}t G\"ottingen, G\"ottingen, Germany}
\affiliation{Institut f{\"u}r Physik, Universit{\"a}t Mainz, Mainz, Germany}
\affiliation{Ludwig-Maximilians-Universit{\"a}t M{\"u}nchen, M{\"u}nchen, Germany}
\affiliation{Fachbereich Physik, Bergische Universit{\"a}t Wuppertal, Wuppertal, Germany}
\affiliation{Panjab University, Chandigarh, India}
\affiliation{Delhi University, Delhi, India}
\affiliation{Tata Institute of Fundamental Research, Mumbai, India}
\affiliation{University College Dublin, Dublin, Ireland}
\affiliation{Korea Detector Laboratory, Korea University, Seoul, Korea}
\affiliation{CINVESTAV, Mexico City, Mexico}
\affiliation{Nikhef, Science Park, Amsterdam, the Netherlands}
\affiliation{Radboud University Nijmegen, Nijmegen, the Netherlands and Nikhef, Science Park, Amsterdam, the Netherlands}
\affiliation{Joint Institute for Nuclear Research, Dubna, Russia}
\affiliation{Institute for Theoretical and Experimental Physics, Moscow, Russia}
\affiliation{Moscow State University, Moscow, Russia}
\affiliation{Institute for High Energy Physics, Protvino, Russia}
\affiliation{Petersburg Nuclear Physics Institute, St. Petersburg, Russia}
\affiliation{Instituci\'{o} Catalana de Recerca i Estudis Avan\c{c}ats (ICREA) and Institut de F\'{i}sica d'Altes Energies (IFAE), Barcelona, Spain}
\affiliation{Stockholm University, Stockholm and Uppsala University, Uppsala, Sweden}
\affiliation{Lancaster University, Lancaster LA1 4YB, United Kingdom}
\affiliation{Imperial College London, London SW7 2AZ, United Kingdom}
\affiliation{The University of Manchester, Manchester M13 9PL, United Kingdom}
\affiliation{University of Arizona, Tucson, Arizona 85721, USA}
\affiliation{University of California Riverside, Riverside, California 92521, USA}
\affiliation{Florida State University, Tallahassee, Florida 32306, USA}
\affiliation{Fermi National Accelerator Laboratory, Batavia, Illinois 60510, USA}
\affiliation{University of Illinois at Chicago, Chicago, Illinois 60607, USA}
\affiliation{Northern Illinois University, DeKalb, Illinois 60115, USA}
\affiliation{Northwestern University, Evanston, Illinois 60208, USA}
\affiliation{Indiana University, Bloomington, Indiana 47405, USA}
\affiliation{Purdue University Calumet, Hammond, Indiana 46323, USA}
\affiliation{University of Notre Dame, Notre Dame, Indiana 46556, USA}
\affiliation{Iowa State University, Ames, Iowa 50011, USA}
\affiliation{University of Kansas, Lawrence, Kansas 66045, USA}
\affiliation{Kansas State University, Manhattan, Kansas 66506, USA}
\affiliation{Louisiana Tech University, Ruston, Louisiana 71272, USA}
\affiliation{Boston University, Boston, Massachusetts 02215, USA}
\affiliation{Northeastern University, Boston, Massachusetts 02115, USA}
\affiliation{University of Michigan, Ann Arbor, Michigan 48109, USA}
\affiliation{Michigan State University, East Lansing, Michigan 48824, USA}
\affiliation{University of Mississippi, University, Mississippi 38677, USA}
\affiliation{University of Nebraska, Lincoln, Nebraska 68588, USA}
\affiliation{Rutgers University, Piscataway, New Jersey 08855, USA}
\affiliation{Princeton University, Princeton, New Jersey 08544, USA}
\affiliation{State University of New York, Buffalo, New York 14260, USA}
\affiliation{Columbia University, New York, New York 10027, USA}
\affiliation{University of Rochester, Rochester, New York 14627, USA}
\affiliation{State University of New York, Stony Brook, New York 11794, USA}
\affiliation{Brookhaven National Laboratory, Upton, New York 11973, USA}
\affiliation{Langston University, Langston, Oklahoma 73050, USA}
\affiliation{University of Oklahoma, Norman, Oklahoma 73019, USA}
\affiliation{Oklahoma State University, Stillwater, Oklahoma 74078, USA}
\affiliation{Brown University, Providence, Rhode Island 02912, USA}
\affiliation{University of Texas, Arlington, Texas 76019, USA}
\affiliation{Southern Methodist University, Dallas, Texas 75275, USA}
\affiliation{Rice University, Houston, Texas 77005, USA}
\affiliation{University of Virginia, Charlottesville, Virginia 22901, USA}
\affiliation{University of Washington, Seattle, Washington 98195, USA}
\author{V.M.~Abazov} \affiliation{Joint Institute for Nuclear Research, Dubna, Russia}
\author{B.~Abbott} \affiliation{University of Oklahoma, Norman, Oklahoma 73019, USA}
\author{B.S.~Acharya} \affiliation{Tata Institute of Fundamental Research, Mumbai, India}
\author{M.~Adams} \affiliation{University of Illinois at Chicago, Chicago, Illinois 60607, USA}
\author{T.~Adams} \affiliation{Florida State University, Tallahassee, Florida 32306, USA}
\author{G.D.~Alexeev} \affiliation{Joint Institute for Nuclear Research, Dubna, Russia}
\author{G.~Alkhazov} \affiliation{Petersburg Nuclear Physics Institute, St. Petersburg, Russia}
\author{A.~Alton$^{a}$} \affiliation{University of Michigan, Ann Arbor, Michigan 48109, USA}
\author{G.~Alverson} \affiliation{Northeastern University, Boston, Massachusetts 02115, USA}
\author{G.A.~Alves} \affiliation{LAFEX, Centro Brasileiro de Pesquisas F{\'\i}sicas, Rio de Janeiro, Brazil}
% ...
%
%
\collaboration{The D0 Collaboration\footnote{with visitors from
$^{a}$Augustana College, Sioux Falls, SD, USA,
$^{b}$The University of Liverpool, Liverpool, UK,
$^{c}$SLAC, Menlo Park, CA, USA,
$^{d}$University College London, London, UK,
$^{e}$Centro de Investigacion en Computacion - IPN, Mexico City, Mexico,
$^{f}$ECFM, Universidad Autonoma de Sinaloa, Culiac\'an, Mexico,
and 
$^{g}$Universit{\"a}t Bern, Bern, Switzerland.
$^{\ddag}$Deceased.
}} \noaffiliation
\vskip 0.25cm
\date{\today}

\begin{abstract}
We present a search for associated Higgs boson production in the process
$p\bar{p} \rightarrow W/ZH \rightarrow \ell^\pm \ell'^\pm + X$ in $ee$, $e\mu$, 
and $\mu\mu$ final states. The search is based on data collected by the D0 
experiment at the Fermilab Tevatron Collider at $\sqrt{s} = 1.96$ TeV corresponding to 
5.3~fb$^{-1}$ of integrated luminosity. We require two isolated leptons (electrons or muons) with the same 
electric charge and additional kinematic requirements. No significant excess 
above background is observed, and we set 95\% C.L. observed (expected) upper limits
on ratio of the production cross section to the standard model prediction
of 6.4 (7.3) for a Higgs boson mass of 165~GeV and 13.5 (19.8) for a mass of 115~GeV.

\end{abstract}

\maketitle


\section{\label{sec:intro}INTRODUCTION}

In the standard model (SM), the Higgs boson decays predominantly to a $WW$ pair 
for Higgs boson masses above 135~GeV and, with a moderate branching fraction, 
to a $\tau\tau$ pair for lower masses, both of which decay to leptonic 
final states involving neutrinos. 
Consequently the associated production of a Higgs boson,
$p\bar{p} \rightarrow$
      $W/ZH \rightarrow~\ell^{\pm}~\ell^{\pm} + X$,
which has additional final state particles, provides easily-detected experimental signatures
comprising two leptons of the same electric charge.
This requirement rejects SM processes with
oppositely charged dileptons that occur with high production rates such as 
$Z/\gamma^*$, $WW$, and $t\bar{t}$. Therefore, the like charge signature from associated vector boson-Higgs 
production has an advantage over direct Higgs production,
$p\bar{p} \rightarrow H \rightarrow~WW$, where only unlike charged leptons 
are produced in the final state.

The D0 Collaboration previously published a search 
for associated Higgs production with the like charge dilepton signature 
based on approximately 400~pb$^{-1}$ of integrated luminosity in 2006~\cite{prev-version-pub}.
The most recent result from a similar search by the CDF experiment using 4.8~fb$^{-1}$
of integrated luminosity~\cite{cdf-hww} was included in the combination of Tevatron searches in the $H \rightarrow~WW$ 
decay mode in 2010~\cite{tevcomb-hww}. 

In this Article, we present a search for associated Higgs boson production 
with like charged dileptons using 5.3~fb$^{-1}$ of integrated luminosity collected by the D0 detector
during the Tevatron Run II period between 2002 and 2009.
The search combines the three leptonic final states with either electrons or muons: 
$ee$, $\mu\mu$ and $e\mu$.

Background processes for the like charge lepton signatures are diboson production,
$p\bar{p} \rightarrow WZ \rightarrow~\ell\nu\ell\ell$ and
$p\bar{p} \rightarrow ZZ \rightarrow~\ell\ell\ell\ell$. Non-resonant
triple vector boson production ($VVV$, $V = W,Z$) and the production
of $t\bar{t}$ + $V$ are negligible.
There are two types of instrumental backgrounds. The first, ``charge flip",
originates from the misreconstruction of the lepton charge.
For the same lepton flavor channels ($ee$ and $\mu\mu$) this
background arises mainly from the Drell-Yan process,
$p\bar{p} \rightarrow Z/\gamma^*\rightarrow \ell^+\ell^-$. When the
two leptons are of different flavor, this background is negligible.
The second instrumental background is from falsely identified leptons which originate 
from jets and photons converting to electrons in $W$ boson or multijet production.
Although these instrumental effects occur at low rate, the associated backgrounds
make sizeable contributions to the dilepton selection due to the large production 
cross sections of the underlying physics processes.

The Higgs boson signal contains multiple neutrinos in the final state, 
hence a complete reconstruction of the Higgs boson mass is not
possible. A Higgs boson signal would appear as an excess of events with like charged leptons
with kinematic properties consistent with $VH$ decay, including 
missing energy from the neutrinos in the leptonic decays of the vector bosons,
as well as additional final state objects, mainly jets, from the decay of the third vector boson. 
A multivariate technique is
employed to provide maximum separation between signal and background
processes based on various kinematic variables. In the absence of an excess over 
the expected number of events from background processes, upper cross section 
limits are set.

\section{\label{sec:selection}EVENT SELECTION}

Events are preselected by identifying at least two leptons. 
The selection of the electrons and muons include kinematic requirements
as well as criteria on their quality.  
The final event selection is performed with a multivariate discriminant, 
which will be discussed later in this Article.

The D0 detector is composed of a central tracking system with a 
silicon microstrip tracker (SMT) and a central fiber tracker (CFT)
embedded within a 2 T solenoidal magnet, preshower detectors, 
a uranium/liquid-argon calorimeter with electromagnetic (EM) and 
hadronic sections, and a muon spectrometer with drift tubes, scintillation counters
and toroidal magnets~\cite{bib:d0det}.  The detector was upgraded in spring
2006 to include, among others, an additional inner layer of silicon
microstrip tracking~\cite{bib:layer0}.  A lepton is identified by the presence of a
track, and its electrical charge is determined by the direction of the track
curvature in the magnetic field. Events with like charged dileptons are 
retained for the analysis while those with unlike charged dileptons are
used to validate the event reconstruction and simulation.

Electrons are characterized by their interaction in the EM calorimeter
and are required to match a track. The energy is measured in the EM and 
the first hadronic layers of the calorimeter within a simple cone of radius
$\mathcal{R} = \sqrt{(\Delta\eta)^2 + (\Delta\phi)^2}$ = 0.2, where $\eta$ and $\phi$ are
the pseudorapidity~\cite{bib:eta} and the azimuthal angle, respectively. 
The electron cluster must satisfy:
	calorimeter isolation fraction, 
	$f_{\rm iso} = [E_{\rm tot}(\mathcal{R}<0.4) - E_{\rm EM}(\mathcal{R}<0.2)]/E_{\rm EM}(\mathcal{R}<0.2)$, 
	less than 0.2, where $E_{\rm tot}$ is the total energy in the isolation cone of radius $\mathcal{R}$ = 0.4 
	and $E_{\rm EM}$ is the EM energy in a cone of radius $\mathcal{R}$ = 0.2;
	EM fraction, $f_{\rm EM} = E_{\rm EM} / E_{\rm tot}$, greater than 0.9, where both energies are measured within 
	the cone of $\mathcal{R}$ = 0.2;
	and the ratio of electron cluster energy to track momentum, $E/p$, between 0.5 and 3.0.
	In addition, the electron candidates are required to have an
 	eight-variable likelihood ($\mathcal{L}_{e}$) greater than 0.85, where the likelihood is calculated from
	$f_{\rm iso}$; $f_{\rm EM}$; $E/p$; number of tracks within the isolation cone; scalar sum of the tracks'
	transverse momenta, $p_{T}$,
	within $0.05 < \mathcal{R} < 0.4$ of the electron track; track-cluster match probability computed
	from the spatial separation and the expected resolution; track distance to the primary vertex
	at closest approach ($dca$); and, lastly, covariance matrices built from 
	the energy depositions in various layers of the calorimeter to represent the longitudinal 
	and lateral shower development.
A loose quality electron is defined by relaxing the requirement on the likelihood 
to $\mathcal{L}_{e} >$ 0.2, and down to 0.01 or 0.0 depending on the purpose, and by
removing the $E/p$ requirement to estimate the lepton fake 
rate and model the backgrounds from $W$ boson and multijet production.

Muons are identified by the presence of at least one track segment reconstructed 
in the muon spectrometer which is spatially consistent with a track in the central detector,
where the momentum and charge are measured by the curvature of this track.
The muon candidate must pass cosmic ray veto timing criteria, be outside a cone of radius $\mathcal{R} < 0.1$
from any jet of particles present in the event, and must not share its track with an electron candidate 
satisfying the calorimeter isolation and EM fraction requirements described above. 
Muon isolation is imposed with two isolation variables defined as the scalar sums of 
the transverse energy, $\Sigma E_{T}^{\rm calo}$, in the calorimeter in an annulus with radius $0.1 < \mathcal{R} < 0.4$, 
and of the momenta of tracks, $\Sigma p_{T}^{\rm trk}$, around the muon candidate within a radius of $\mathcal{R} = 0.5$.
Each of the two isolation variables must be less than 2.5~GeV.
Relaxed isolation criteria define loose muon quality where the track isolation upper bound is raised
to $\Sigma p_{T}^{\rm trk} < 10$~GeV and the calorimeter isolation is ignored.

Both leptons are restricted to the central region within the CFT coverage,
$|\eta_{e}| < 1.1$ for electrons and $|\eta_{\mu}| < 1.8$ for muons, and have transverse momenta 
in the range 15~GeV $< p_{T} <$ 200~GeV, together constraining the dilepton invariant mass 
to 15~GeV $< M(\ell,\ell) <$ 250~GeV. The upper thresholds were chosen to eliminate imprecise
charge measurements in the tracking system.
The signal loss due to the $|\eta|$ requirements is non-negligible, however, 
the reduction of instrumental backgrounds leads to an overall gain in the signal-to-background ratio.
The two lepton tracks are required to have at least one hit in the SMT. The longitudinal (transverse) 
distance separating the point of closest approach between the tracks and the primary $p\bar{p}$ interaction vertex
must be less than 1.0 (0.01)~cm. The best interaction vertex, 
which is selected based on the number of associated tracks and their transverse momenta,
must be within $|z| <$ 60~cm, where $z$ is the longitudinal coordinate measured from the detector center.
Additional leptons that satisfy the same lepton and track quality requirements 
are allowed up to $|\eta_{e}| < 2.5$ and $|\eta_{\mu}| < 2.0$. 
No upper $p_T$ bound is imposed for these leptons.

The hadronic decays of vector bosons and partons are identified as jets of particles,
and are reconstructed by clustering the calorimeter energy deposition within
a cone of radius $\mathcal{R}$ = 0.5~\cite{bib:jet}. 
Jets are required to have $p_{T} >$ 15~GeV and $|\eta| <$ 2.5. 
At least two tracks originating from the primary vertex must be found within the jet cone.
Momentum imbalance in the transverse plane of the event implies the presence of the neutrino; 
the missing transverse energy, \met, is reconstructed as the negative vector sum of the energies
in the calorimeter towers and 
the muon momenta subtracting the calorimeter energy deposition due to muons.

\section{\label{sec:simulation}EVENT SIMULATION}

Associated production of the Higgs boson and the diboson background
processes are modeled by Monte Carlo (MC) event generators and the detector response is provided 
by the {\sc geant} based~\cite{bib:geant} simulation of the D0 detector. 
The effect of additional $p\bar{p}$ interactions is reproduced 
by overlaying data events taken from random collisions onto the MC generated processes.
The signal and diboson processes are simulated by the {\sc pythia} event generator~\cite{bib:pythia} 
using CTEQ6L1~\cite{bib:cteq} parton distribution functions (PDFs).
The signal event samples are generated for different Higgs boson masses between 
115~GeV and 200~GeV with 5~GeV intervals, and normalized to the next-to-next-to-leading 
order (NNLO) cross sections calculated using MSTW2008 PDFs~\cite{bib:vh-xs}. 
Branching ratios of Higgs boson decays are provided by {\sc hdecay}~\cite{bib:hdecay}.
The diboson cross sections are calculated
at next-to-leading order (NLO) using CTEQ6.1M PDFs~\cite{bib:dibo-xs},
and the transverse momentum of the diboson system is reweighted to 
the prediction from {\sc mc$@$nlo}~\cite{bib:mcatnlo}.
Production of $W$ bosons with additional emission of partons is simulated with the 
{\sc alpgen} event generator~\cite{bib:alpgen}, which implements matrix element calculations,
and the hadronization process is simulated by {\sc pythia}. The NNLO total cross sections
calculated using MRST2004 PDFs~\cite{bib:v-xs} are used to normalize the sum of the $W$ production processes with
0 through 5 associated light partons and the $W$ production processes with heavy flavour quarks ($b\bar{b}$ or $c\bar{c}$).

The unlike charge dilepton events are used to study the event reconstruction
and selection as well as instrumental effects. The events include physics processes 
which do not contribute to the like charge selection: $Z/\gamma^*$, $t\bar{t}$ and $WW$ production.
These processes are simulated using the {\sc pythia} generator, as described above. 
The $Z/\gamma^*$ total cross sections are determined at NNLO, the $WW$ cross section at NLO,
and the $t\bar{t}$ cross section is obtained at approximate NNLO~\cite{bib:tt-xs}.

The overall normalization of the MC samples are initially obtained from the 
corresponding theoretical cross sections and the integrated luminosity recorded by the D0 detector. 
The kinematic dependences of the efficiencies of the triggers, lepton identification and vertex selection 
as well as the mismodeling of instantaneous luminosity profile and beam spot position are corrected. 
Additional scale factors specific to this analysis are obtained by comparing the yields 
of the unlike charge dilepton events within the $Z$ resonance in the 
dilepton invariant mass between 70~GeV and 110~GeV. 
The normalization factors include residual effects due to the efficiencies of triggers, 
lepton identification, track quality requirements, vertex selection, 
as well as the measurement of the integrated luminosity. 
The efficiencies specific to the electrons and the muons are measured with
dielectron and dimuon events respectively, while the effects that are common
to the two lepton types have the same multiplicative contributions
to the total normalization factors in both samples.

\section{\label{sec:backgrounds}INSTRUMENTAL BACKGROUNDS}

The two largest instrumental backgrounds, charge mismeasurement and
lepton misidentification, are estimated from data,
either by measuring their rate using control samples enriched with the particular background processes 
or by performing a fit to the kinematic distribution to predict their fraction in the analysis sample.
The separate contributions to the background from charge flip and $W$ and 
multijet production are discussed below.

\subsection{\label{sec:chargeflip}Charge Flip}

The charge flip background, created by mismeasurement of the
charge of one of the leptons, mostly originates from the $Z/\gamma^*$
process. This occurs when the curvature of a high \PeTe~track
is not correctly measured, or when additional hits from other
charged particles and noise are present near the track. For electrons,
conversion of photons from bremsstrahlung radiation is also estimated 
as part of the charge flip background,
although the charge may be correctly measured.

Contributions from charge flip in the $\mu\mu$ sample are estimated
using two uncorrelated measurements of the lepton charge; the first one is the
measurement of the track curvature in the central tracker, and the second
measurement, called ``local'' muon charge, is measured by the muon
spectrometer. The like charge dimuon samples are categorised into three types;
events in which the two charge measurements give the same answer for both leptons (AA), 
agree for one lepton and disagree for another one (AD), or disagree for both leptons (DD). 
The number of events in these three categories depends on the actual number of 
like charge $\mu\mu$ events that originate from charge mismeasurements, $N^{\rm flip}$, 
and those that originate from true like charge processes such as dibosons, $N^{\rm true}$,
as well as on
the efficiency of the local charge measurement, $\epsilon_{loc}$,
which gives the probabilities of the AA/AD/DD configurations, $P$, for each case.
\begin{eqnarray}
\label{cf_eq}
N_{\rm AA}&=&P^{\rm true}_{\rm AA}(\epsilon_{loc}) \cdot N^{\rm true} + P^{\rm flip}_{\rm AA}(\epsilon_{loc}) \cdot N^{\rm flip},\nonumber\\*
N_{\rm AD}&=&P^{\rm true}_{\rm AD}(\epsilon_{loc}) \cdot N^{\rm true} + P^{\rm flip}_{\rm AD}(\epsilon_{loc}) \cdot N^{\rm flip},\\*
N_{\rm DD}&=&P^{\rm true}_{\rm DD}(\epsilon_{loc}) \cdot N^{\rm true} + P^{\rm flip}_{\rm DD} (\epsilon_{loc}) \cdot N^{\rm flip}\nonumber .
\end{eqnarray}
For a true like charge event to appear in the AA configuration, both leptons must have 
the correct local charge measurements, giving $P^{\rm true}_{\rm AA} = \epsilon_{loc} \cdot \epsilon_{loc}$.
For a charge flip event where the central tracker mismeasures the charge of one lepton, 
the local charge must also be incorrect, hence $P^{\rm flip}_{\rm AA} = \epsilon_{loc} (1 - \epsilon_{loc})$.
The parameterization of $\epsilon_{loc}$ is measured using unlike charge 
$Z \rightarrow~\mu\mu$ events as a function of 1/\PeTe~of the muon. 
The fraction of the charge flip events within the like charge $\mu\mu$ sample,
\begin{center}
	$f^{\rm flip} = N^{\rm flip} / (N^{\rm flip} + N^{\rm true})$,
\end{center}
is determined by solving the over-constrained equations yielding 
$f^{\rm flip} = 0.95 \pm 0.14^{1}$\footnotetext[1]{The positive uncertainty is constrained by the total background yield at later stage to obtain the final result.}.
The ratio of the predicted number of charge flip events to the number of unlike charge events 
gives a charge mismeasurement rate in the dimuon sample of approximately $10^{-3}$.

A different method is employed to estimate the charge flip contribution to the $ee$ background,
for which a second charge measurement does not exist.
The charge flip event rate is measured with data using a control region enriched
in $Z \rightarrow~e^+e^-$ events, defined by a dielectron invariant mass
reconstructed with precisely measured energies in the calorimeter,
85~GeV $< M(e,e) <$ 100~GeV 
and an azimuthal separation between the leptons of
$\Delta\phi(e,e)$ $>$ 2.8.
The charge flip rate corresponds to the like charge fraction in the $Z$ boson enriched region, 
removing contributions from known sources of true like charged dileptons such as
diboson, $W$ boson and multijet production. 
The measured charge flip event rate in the control region in data after the dielectron selection 
is $(8.5 \pm 1.4) \times 10^{-4}$. 
The rate is then used to scale the unlike charge distribution of the data outside the control region 
to obtain the charge flip prediction for the events selected for analysis.

The contribution of charge flips to the $e\mu$ selection is negligible as
the dominant $Z/\gamma^{*}$ production must decay via a $\tau$ lepton pair 
and is suppressed by the branching fraction of the $\tau$ lepton into an electron or a muon.
In addition, leptons from $\tau$ decays have a lower \PeTe~spectrum, 
hence the average charge flip rate is smaller.

The kinematic distributions of the charge flip events are modeled by unlike charge data,
with corrections applied for the effects of charge mismeasurement derived using MC leptons. 
The corrections include the dependence of the charge flip rate on the lepton $p_{T}$ and $\eta$ 
in the detector, and the resolution of the muon momentum measured
from the track curvature, parameterized as a function of 1/\PeTe.

\subsection{\label{sec:wjet}\boldsymbol{$W$} Production}

The production of $W$ bosons contributes to the like charge dilepton background 
when there is a ``fake" lepton which can originate from jets 
or when there are photon conversions to electrons. 
The contribution from these backgrounds is estimated using MC simulations, taking
into account corrections derived from data for the electron misidentification rate and photon conversions.

The corrections to these two subcategories in the MC samples are obtained 
in two steps using template fits to the kinematic distributions 
in a $W$ boson enriched region in the $ee$ sample, which
requires one of the electrons to have loose quality, defined by the $\mathcal{L}_{e}$ variable,
and to fail the tight selection used for the analysis.
The first step determines the normalization of the $W$ boson background
as well as the charge flip and multijet contributions in the control region
using a two-dimensional distribution defined by $M(e,e)$ and \met.  
The second step determines the fractions of the two contributions within the $W$ boson sample, jets and photons, 
while keeping the overall normalization of the total sum derived in the first step.
The presence of a hit in the first layer of the SMT detector is used
to discriminate between the two contributions; 
the photon conversion rate scales as the amount of material the photon traverses,
hence a sizeable fraction of electrons from photon conversion are expected 
to have no hit in the first SMT layer.
To increase the purity of the $W$ process in the control sample,
we require \met~$>$ 25~GeV and $M_{T}(e,\met) >$ 40~GeV, where 
$M_{T}(e,\met) = \sqrt{2 \cdot p_{Te} \cdot \met (1 - \cos \Delta\phi)}$,
where $\Delta\phi$ is the angle in the azimuthal plane between the electron and the direction of \met,
and $e$ is the electron which gives the maximum value of $M_T$. 
The requirement on the electron likelihood is relaxed to 
a minimal value of $\mathcal{L}_{e} >$ 0 for the first step
to maximize the number of $W$ boson events in the control region,
and raised to 0.01 for the second stage to increase the $W$ boson + photon fraction.
The normalization factor obtained in the first stage is $1.2$,
and the correction factors for $W$ + jet and $W$ + photon contributions 
obtained in the second step are $0.95$ and $1.01$, respectively.
The total systematic uncertainty on the $W$ boson background prediction are discussed later.

The efficiency for a jet identified as a loose electron, to pass as a tight electron
is determined from a dedicated dijet data sample where one of the jets is identified as
a loose quality electron candidate. The events are required to contain
only one lepton candidate, back-to-back with a jet, $\Delta\phi(e,{\rm jet}) >$ 2.5,
have low transverse mass, $M_{T}(e,\met) <$ 40~GeV, and invariant mass with
any track lower than the $Z$ resonance, $M(e,{\rm trk}) <$ 60~GeV, to suppress 
contamination from true electrons in $W/Z$ decays. 
Additional physics contamination is estimated by comparing the shape of the
$\mathcal{L}_{e}$ variable for the electron candidates in the fake data sample 
to that in MC dijet events.
The measured fake rate, parameterized in $p_{T}$ and $\eta$,
is then used to scale the $W$ + jet MC sample
selected with the tight-loose requirement to the tight-tight region.
No additional correction is made to the $W$ + photon content of the 
MC sample as the electrons from photon conversions are correctly 
modeled by the simulation. 

\subsection{\label{sec:chargeflip}Multijet}

In the case of jets misidentified as muons, the multijet background
 contains muons from semileptonic decays of heavy flavor quarks, 
punch-through hadrons in the muon detector, and muons from pion or kaon decays in flight. 
In the case of jets misidentified as electrons, the multijet
background contains electrons from semileptonic heavy flavor decays,
from hadrons misidentified as electrons, and from photon conversions.

Multijet contributions to the $ee$ and $\mu\mu$ samples are estimated
from data events in a control sample containing two loose quality like
charged leptons using the loose-to-tight efficiency of fake leptons,
$\varepsilon_{F}$, as measured in that sample.  Within a sample of pure
multijet events, the fraction of events with no tight lepton ($f_0$) and with
exactly one tight lepton ($f_1$) is
\begin{eqnarray}
f_0    &=&  (1-\varepsilon_{F})[1-\varepsilon_{F}(1-\rho)],\nonumber\\
f_1    &=& 2(1-\varepsilon_{F})\varepsilon_{F}(1-\rho) ,
\end{eqnarray}
where $\rho$ is the correlation coefficient, reflecting that the identification of a lepton 
may be different in events where another lepton has been already identified.
These equations are solved for $\varepsilon_{F}$ and $\rho$.
Then the fraction of multijet events with two leptons passing
the tight selection criteria is given by
\begin{eqnarray}
f_2    &=& \varepsilon_{F}^2(1-\rho) + \rho \varepsilon_{F} .
\end{eqnarray}

The control region is 30~GeV $< M(\ell,\ell) <$ 50~GeV and
$\Delta\phi(\ell,\ell)$ $>$ 2.5, where $\Delta\phi(\ell,\ell)$ is the 
azimuthal angle between the two leptons.  
The correlation coefficient is $0.01 \pm 0.05$ for the $ee$ channel and $0.18 \pm 0.06$ for the $\mu\mu$ channel.  
The efficiencies are $0.169 \pm 0.014$ and $0.091 \pm 0.009$ for the $ee$ and the $\mu\mu$ channels, respectively.
Examination of the variation of
$\varepsilon_{F}$ and $\rho$ as track quality requirements are removed shows
no sign of charge flip contamination in the control region, and the spectrum
of $M_{T}(\ell,\met)$ shows no sign of $W$ boson contamination.  Our ability
to identify contaminations from other background processes (in particular events with charge flips) 
is limited by the statistics
of the data sample in the control region.  As such contaminants would raise the
background estimate, we add a one-sided uncertainty in quadrature on the low
side of the estimate, in amount equal to the statistical uncertainty in the
estimate.

The number of multijet events in the $e\mu$ selection is estimated by
performing a fit of the $\mathcal{L}_{e}$ distribution to templates of true and
fake electron taken from data.  
While the $\mathcal{L}_{e}$ requirement must
be relaxed for this fit, tight $E/p$ requirements for the electrons and tight
muon selection requirements are retained.  The template distribution
for true electrons is obtained from the $ee$ pairs at the $Z$ resonance.  The
template distribution for fake electrons is obtained from like charge data events
in a similar control region as is
used in the $ee$ and $\mu\mu$ case.  The method estimates all contributions
from processes containing fake electrons, including $W$ + jet production,
hence the corresponding process is removed from the $W$ MC sample.

The shapes of the kinematic distributions for multijet backgrounds are modeled
by events in like charge data with loose lepton quality but failing the tight
criteria.  The electron requirement is relaxed to $\mathcal{L}_{e} >$ 0 to obtain
a sufficient number of events.  The lepton quality requirements are inverted for both
leptons in $ee$ and $\mu\mu$ channels. For the $e\mu$ channel, events with 
a tight muon are also used to model the $W$ + jet contribution included in the multijet
rate prediction.  The kinematic dependence due to the inversion of lepton qualities
is corrected using the loose-to-tight efficiency for fake leptons obtained in
the dijet sample and by the method described in the section for $W$ boson
background.  For the $\mu\mu$ selection, the multijet background mostly
comes from the semileptonic decays of heavy flavor quark decays; 
hence the correction derived from the dijet sample is not used for the
$\mu\mu$ channel.

\section{\label{sec:tmva}MULTIVARIATE ANALYSIS}

A multivariate technique is employed to characterize the Higgs boson signal 
and the backgrounds and to achieve maximum separation between them. 
A Boosted-Decision-Tree (BDT) algorithm~\cite{bib:bdt} is used 
to construct a discriminant from kinematic variables
taking into account the variable correlations.
The algorithm is robust against low number of events,
which is particularly beneficial for the like charged dilepton search
where the data samples used to model the instrumental backgrounds are limited.

\begin{figure*}[!]
\begin{center}
\begin{tabular}{cc}
{\scriptsize\bf (a)}& {\scriptsize\bf (b)}\\
\vspace{-1mm}
\includegraphics[scale=0.39]{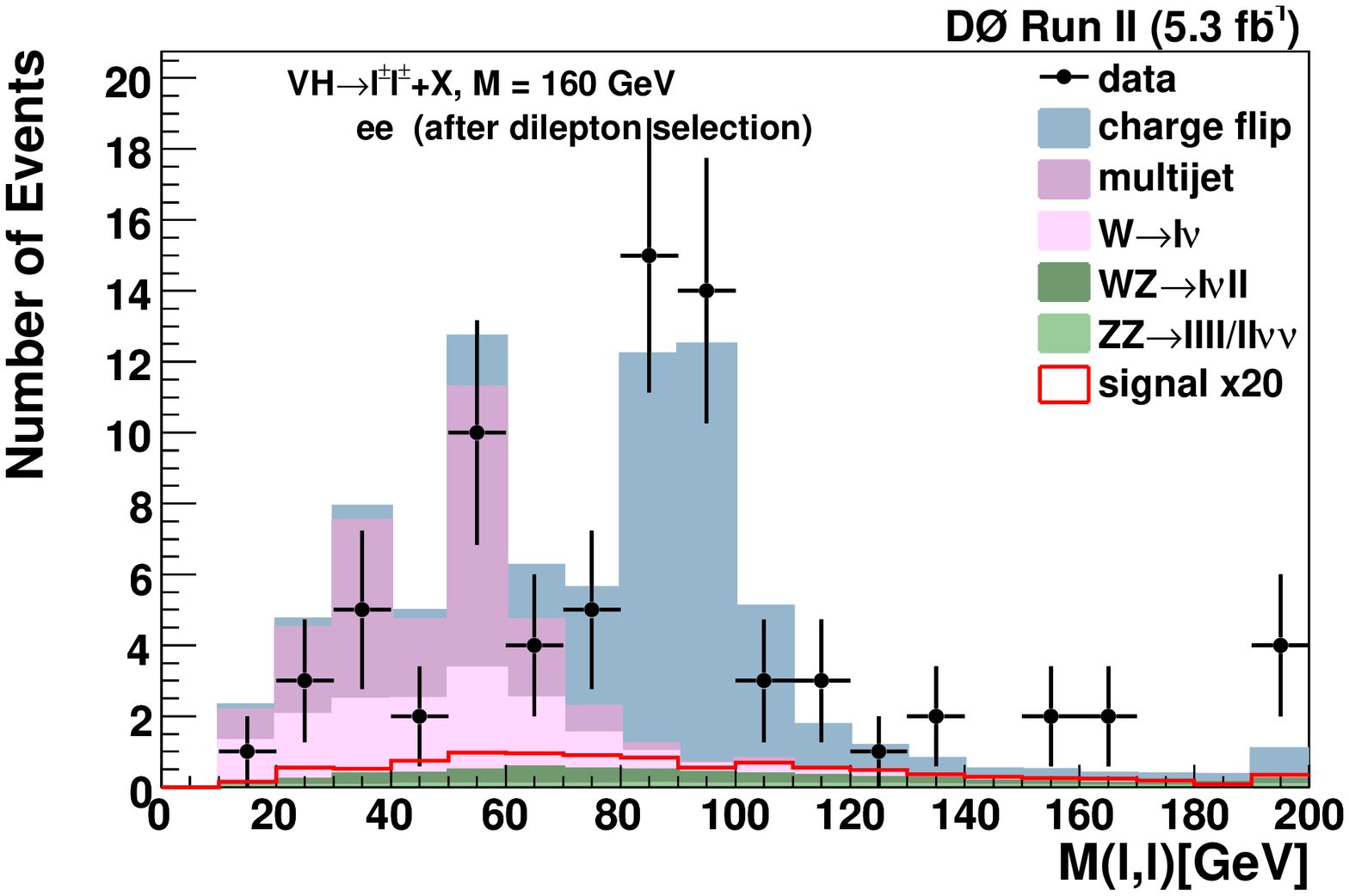}
&\includegraphics[scale=0.39]{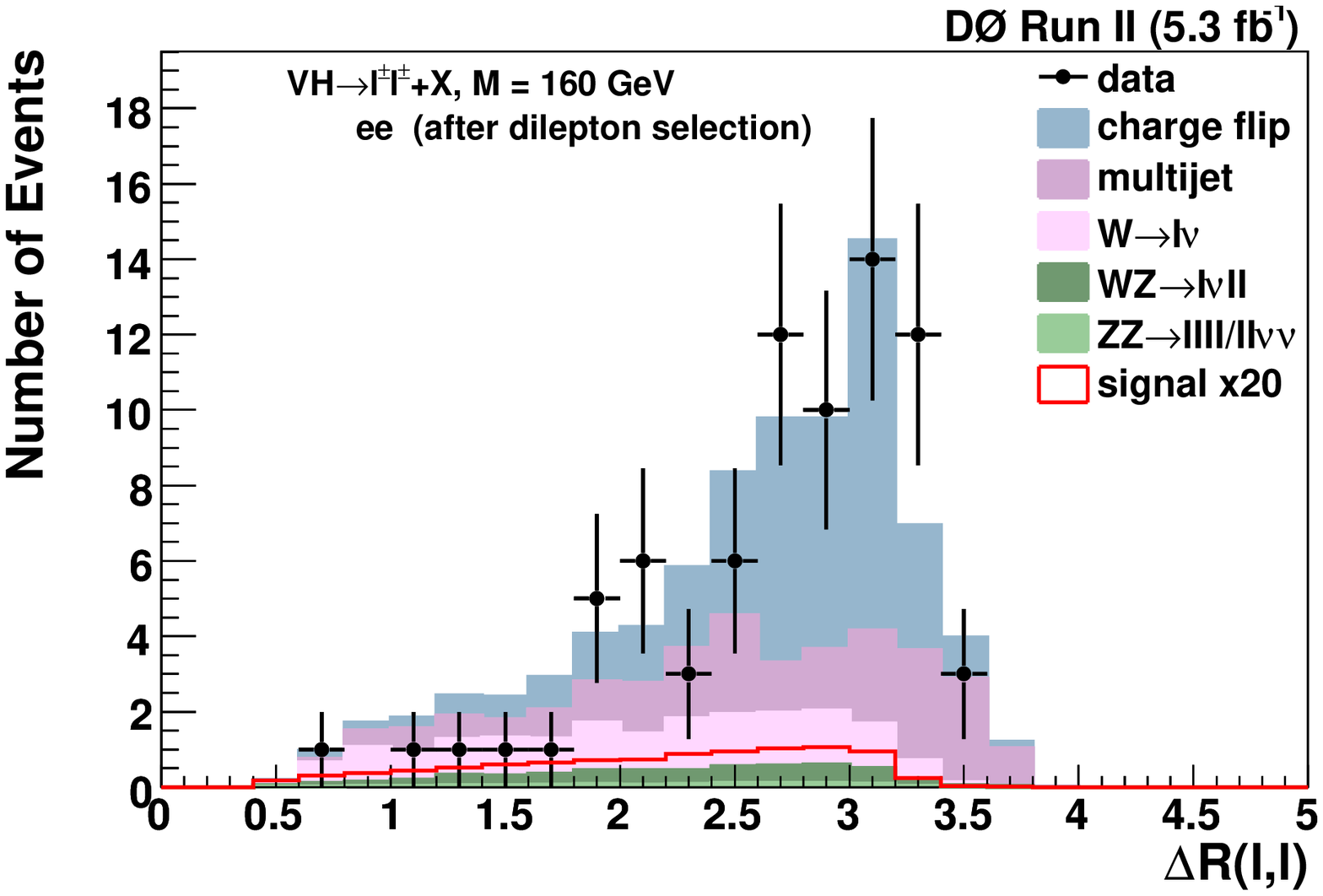}\\
{\scriptsize\bf (c)}& {\scriptsize\bf (d)}\\
\vspace{-1mm}
 \includegraphics[scale=0.39]{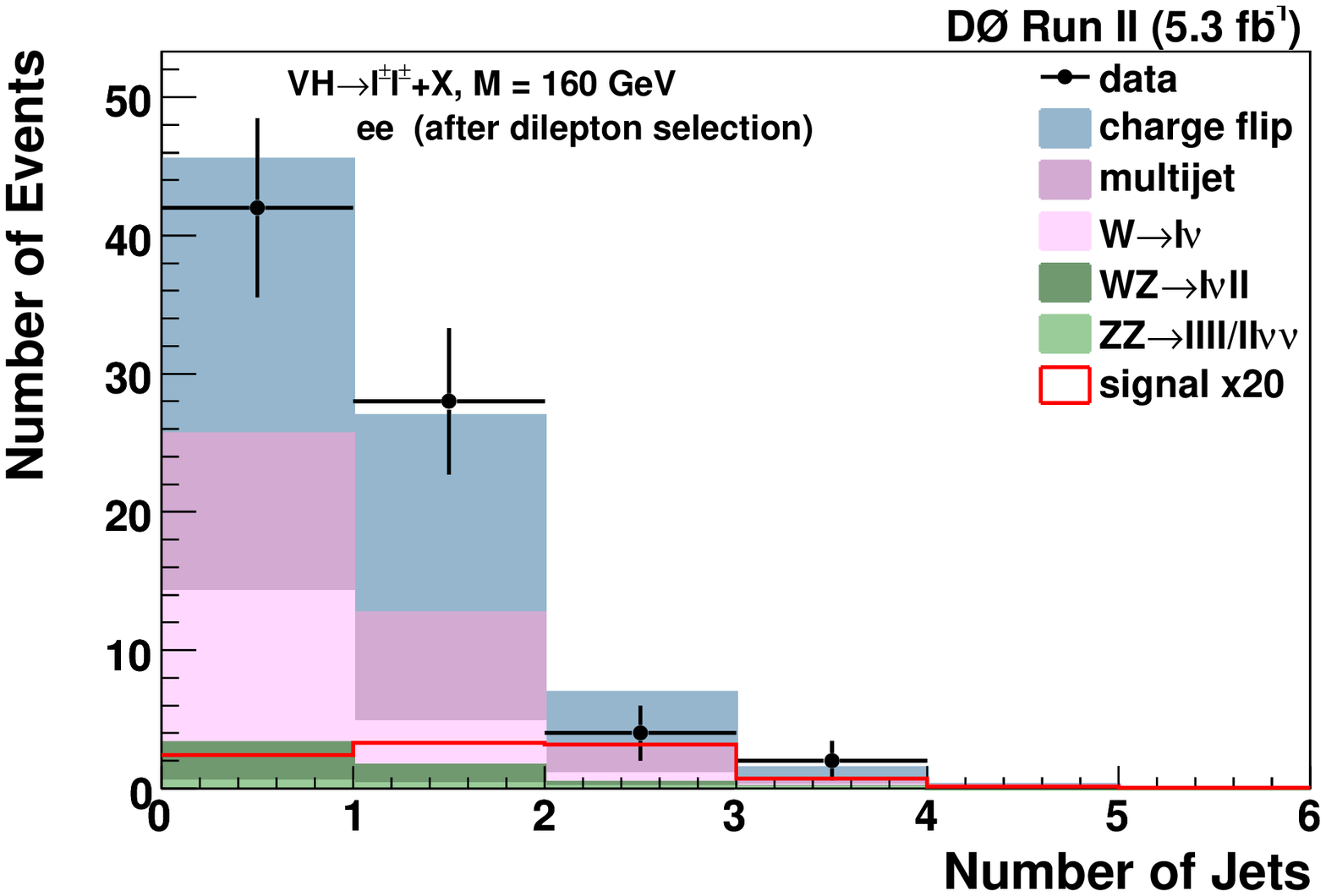}
&\includegraphics[scale=0.39]{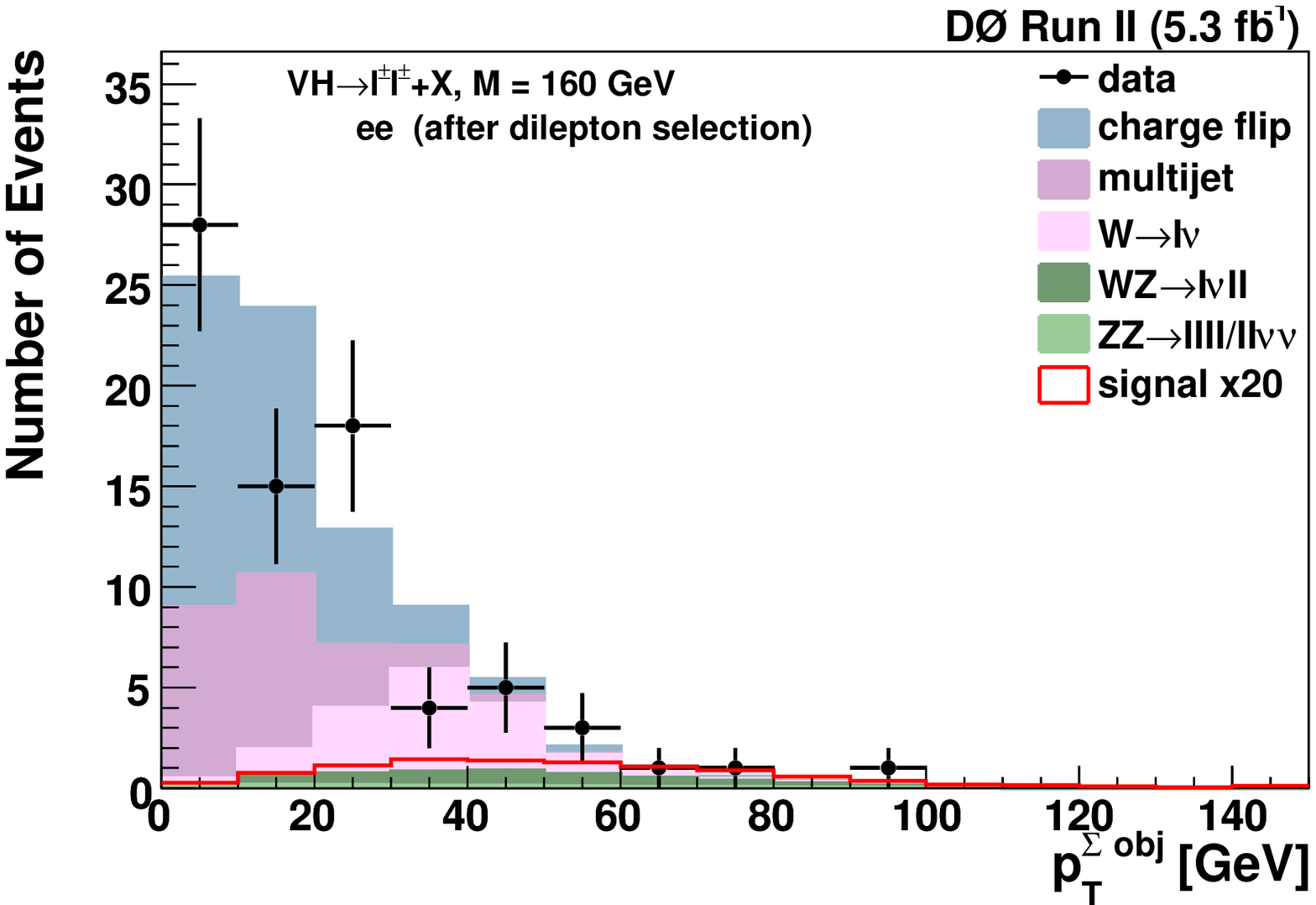}\\
{\scriptsize\bf (e)}& {\scriptsize\bf (f)}\\
\vspace{-1mm}
 \includegraphics[scale=0.39]{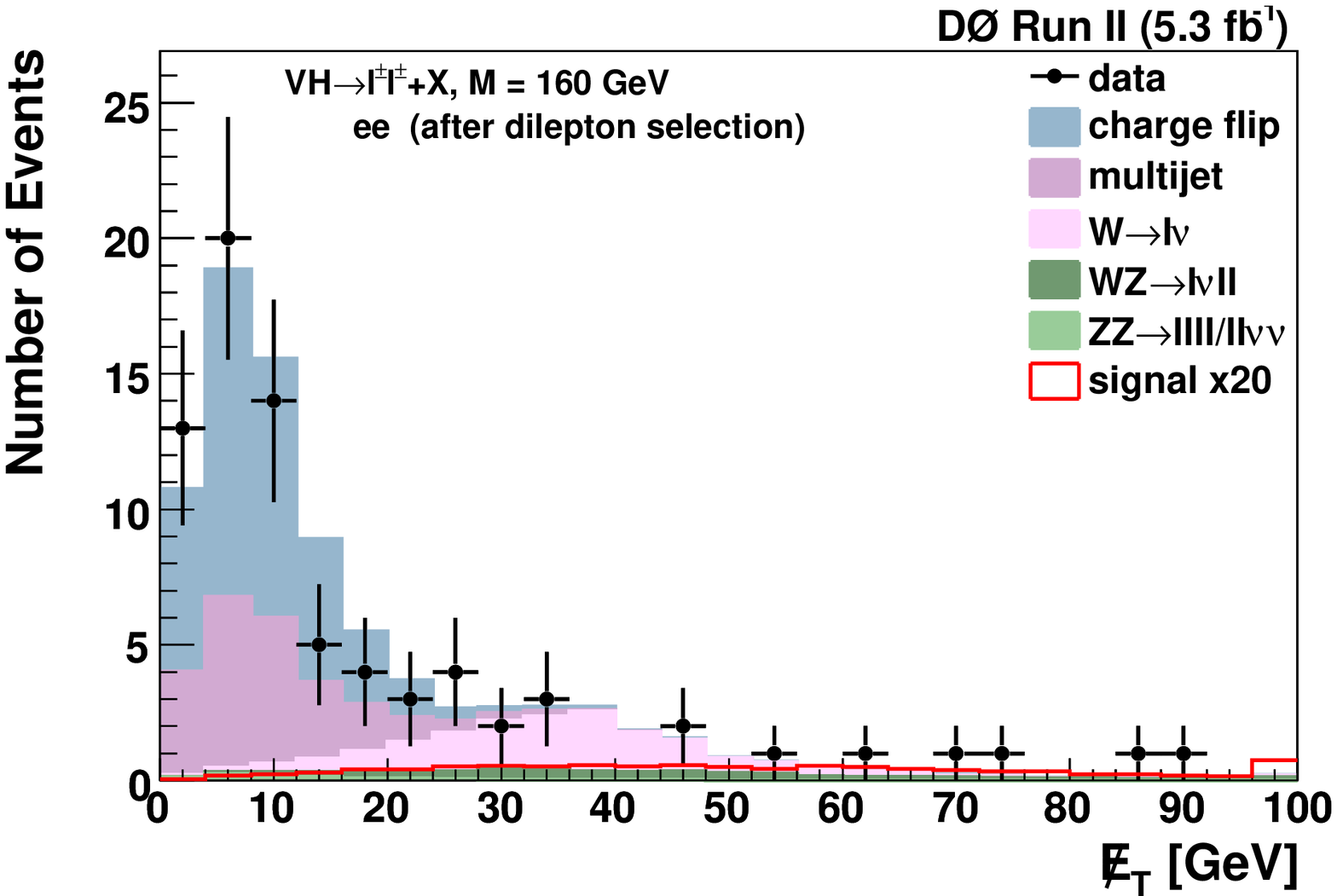}
&\includegraphics[scale=0.39]{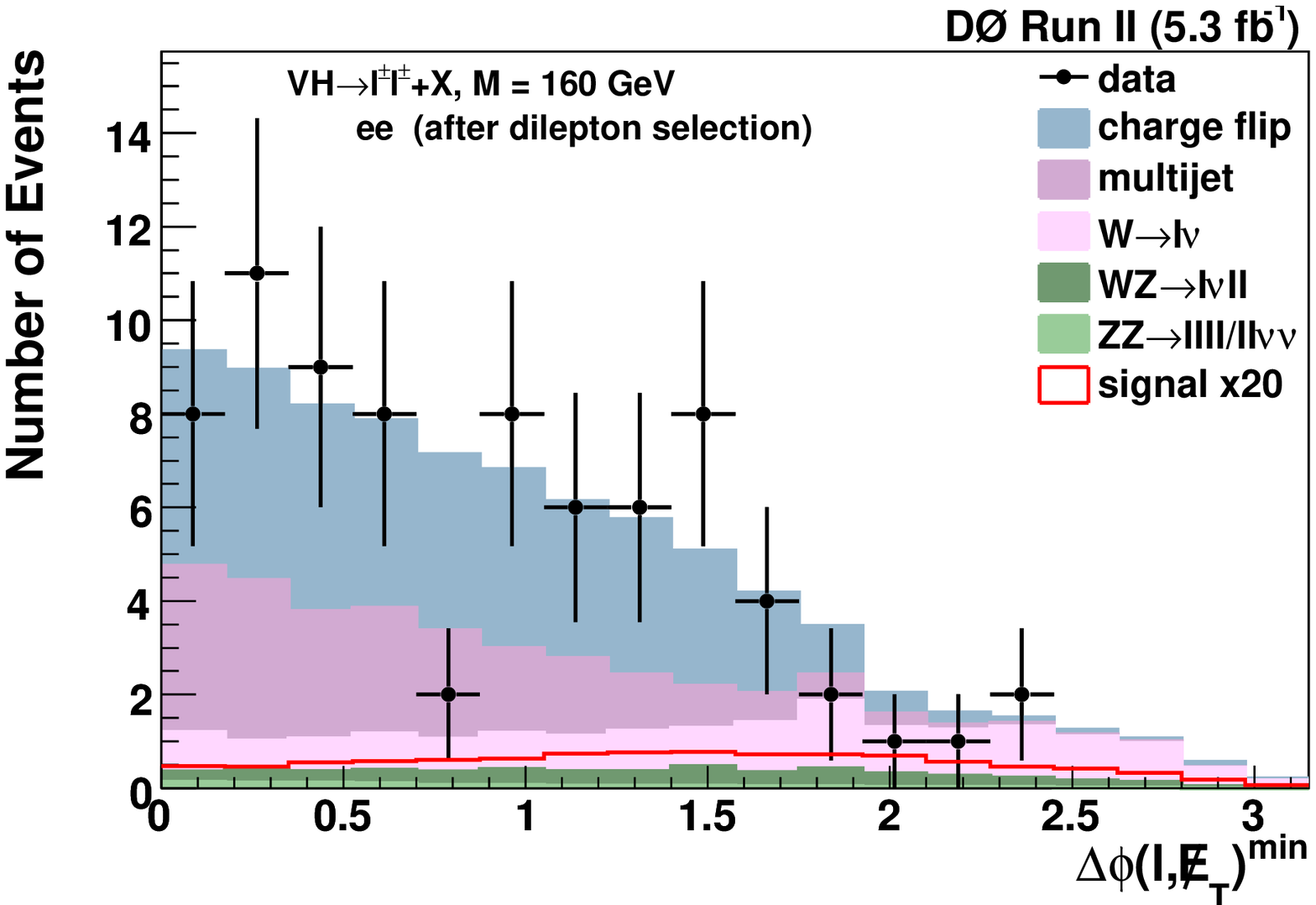}\\
\end{tabular}
\caption{(Color online) The distribution of (a) $M(\ell,\ell)$, (b) $\Delta R(\ell,\ell)$, (c) $N_{\rm jet}$, 
         (d) $p_{T}^{\Sigma {\rm obj}}$, (e) \met, (f) $\Delta\phi(\ell,\met)^{\rm min}$, for the $ee$ channel
  comparing data and predicted backgrounds as well as the Higgs boson signal ($M_{H}$ = 160~GeV)
  expectation after the kinematic selection of like charge dilepton events. }
\label{fig:kine_ee}
\end{center}
\vspace{-2mm}
\end{figure*}

\begin{figure*}[!]
\begin{center}
\begin{tabular}{cc}
{\scriptsize\bf (a)}& {\scriptsize\bf (b)}\\
\vspace{-1mm}
\includegraphics[scale=0.39]{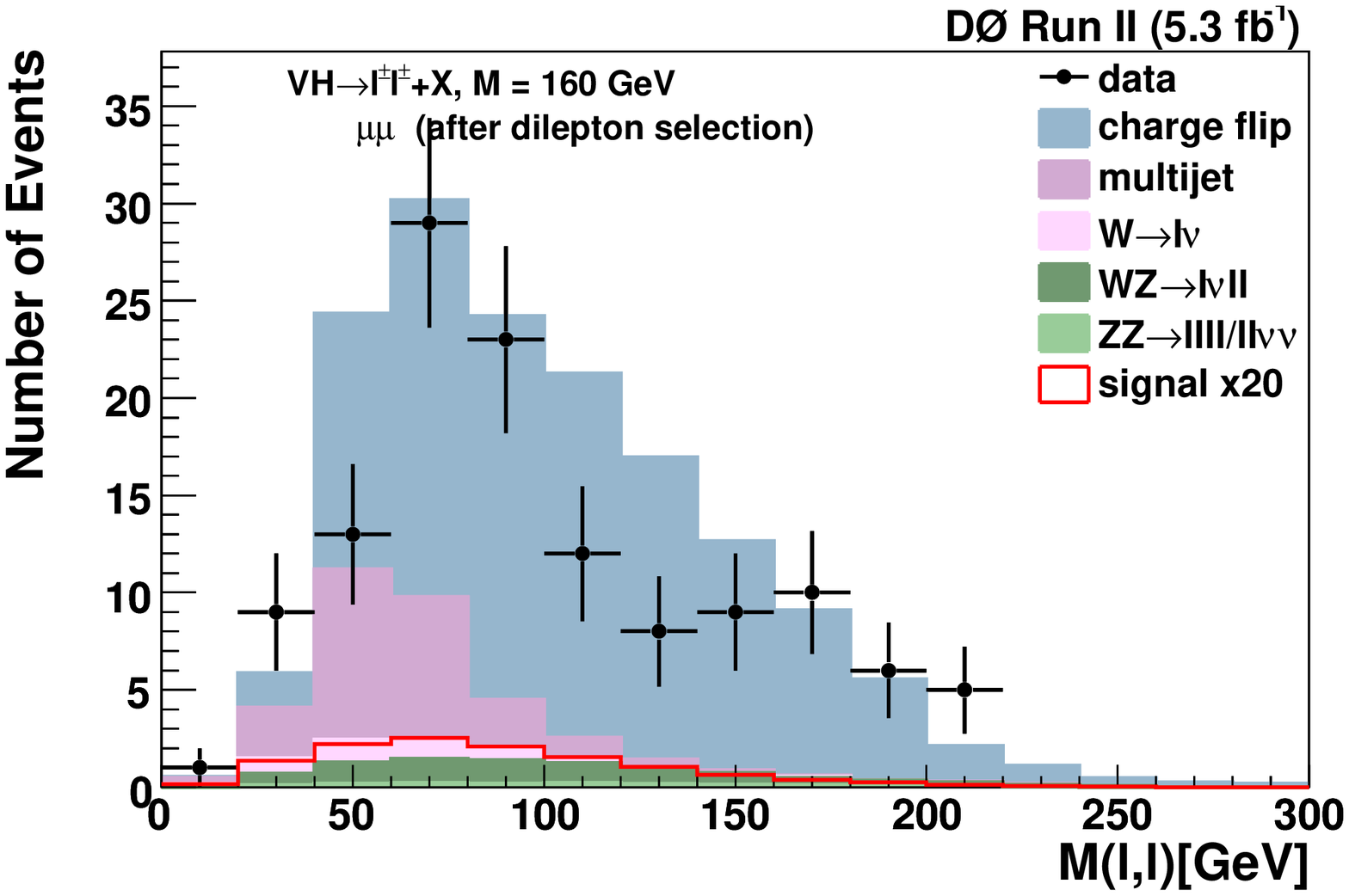}
&\includegraphics[scale=0.39]{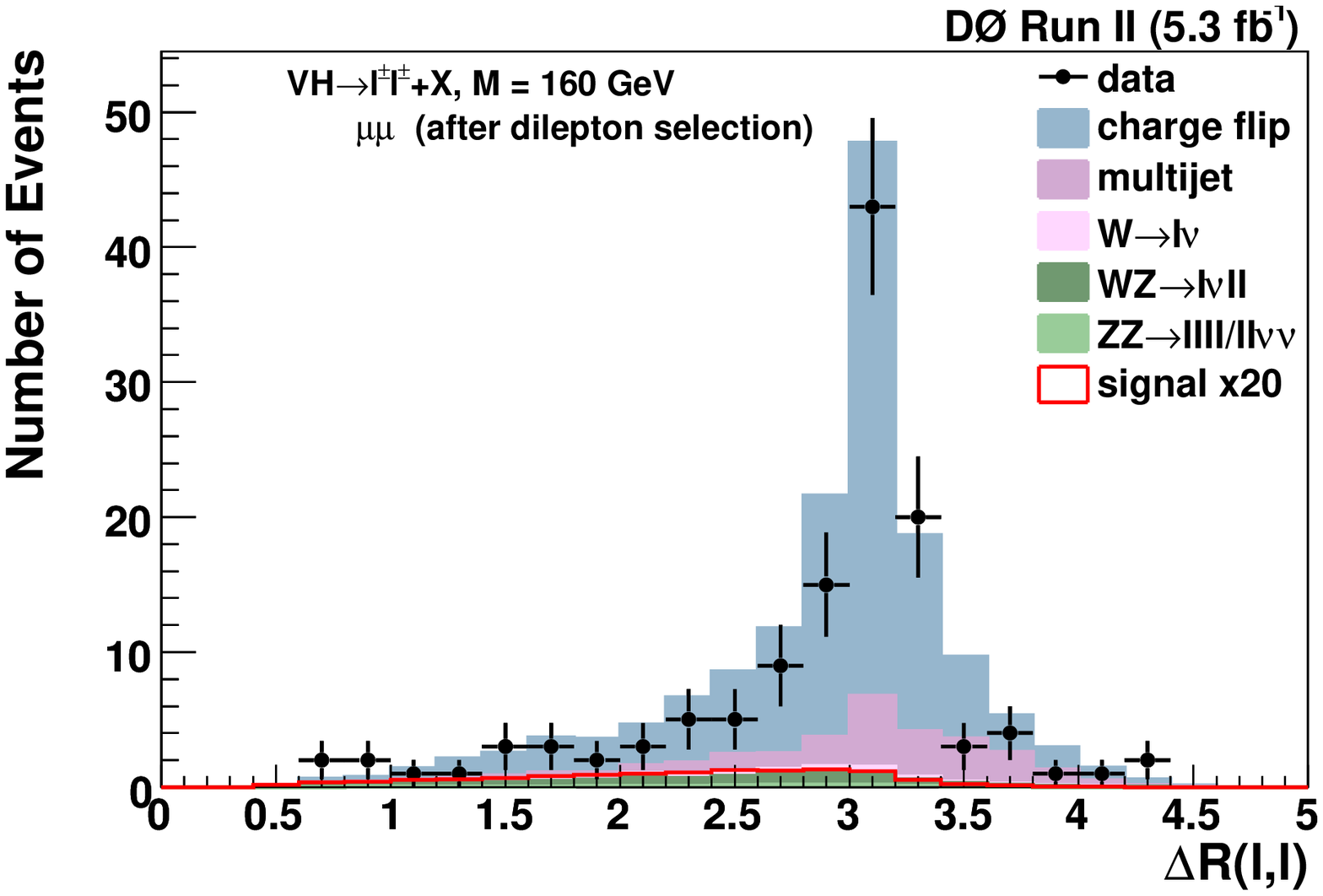}\\
{\scriptsize\bf (c)}& {\scriptsize\bf (d)}\\
\vspace{-1mm}
 \includegraphics[scale=0.39]{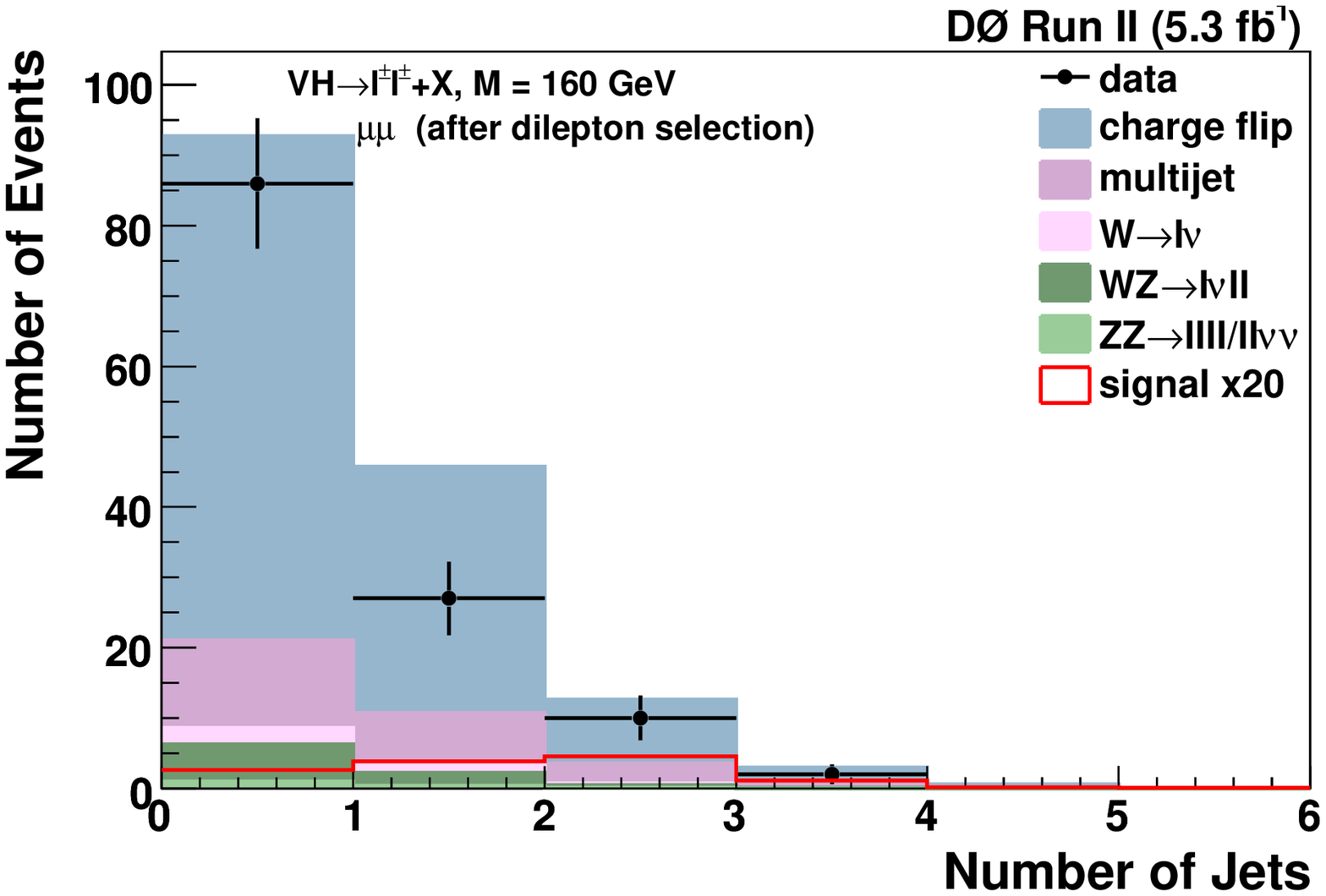}
&\includegraphics[scale=0.39]{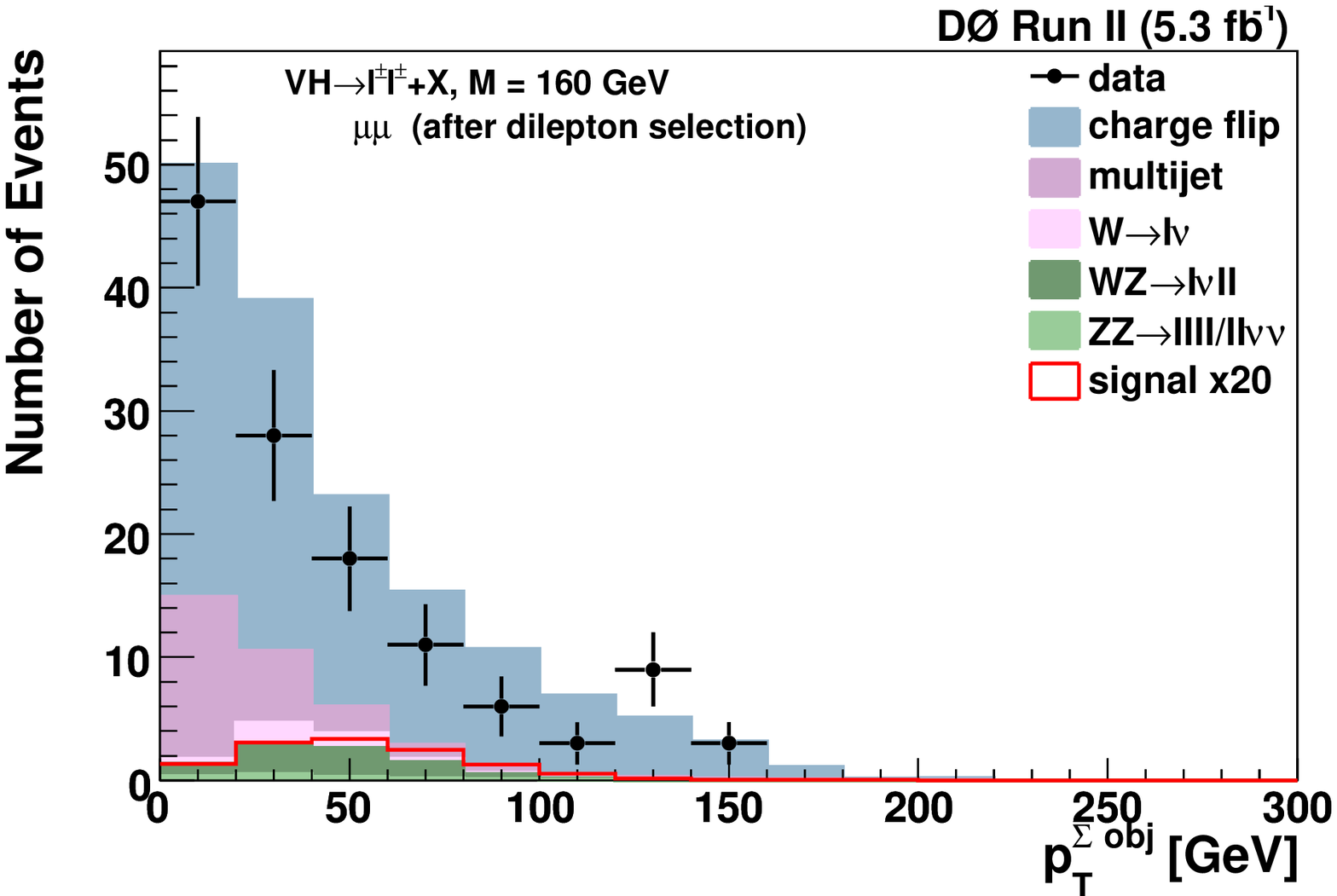}\\
{\scriptsize\bf (e)}& {\scriptsize\bf (f)}\\
\vspace{-1mm}
 \includegraphics[scale=0.39]{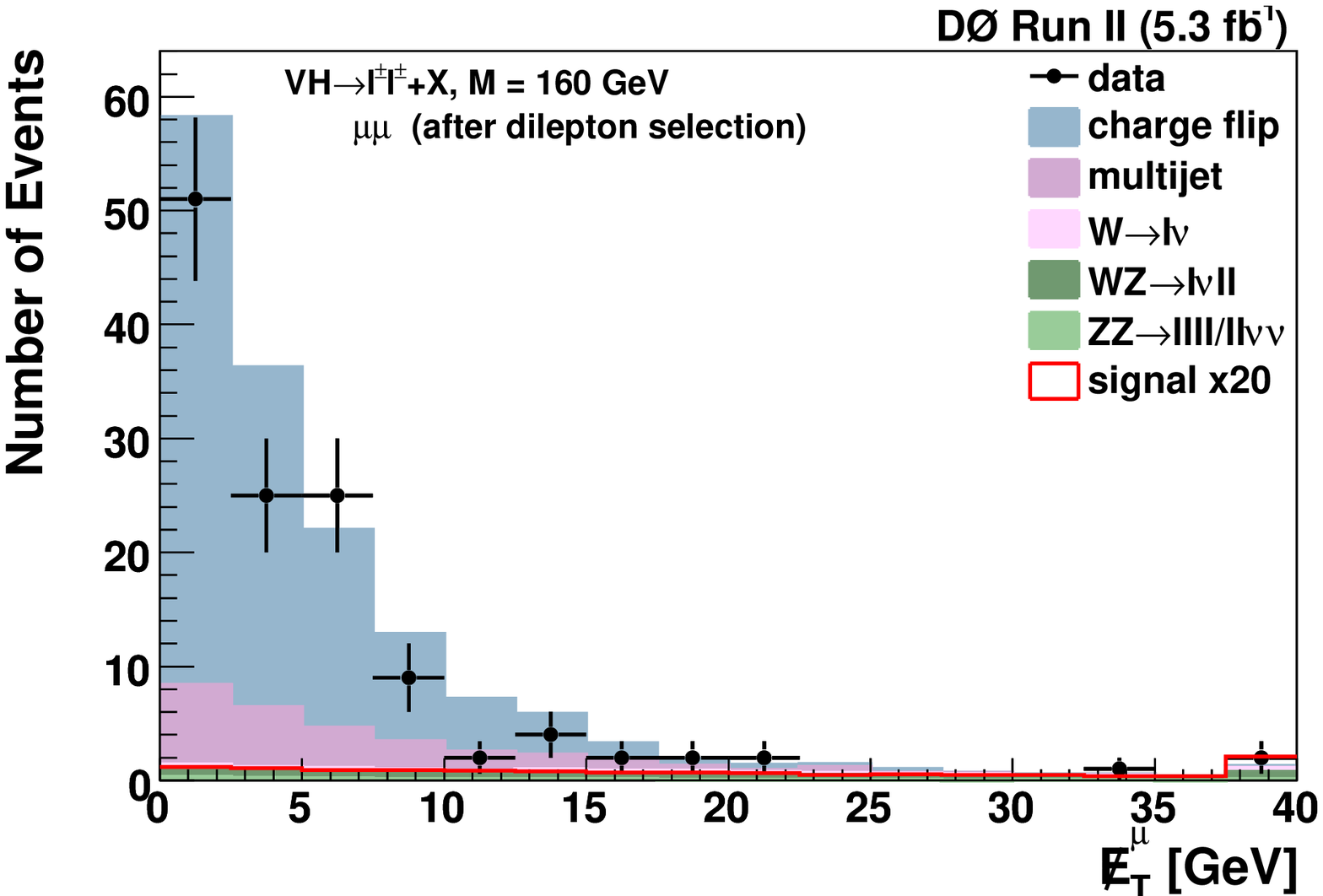}
&\includegraphics[scale=0.39]{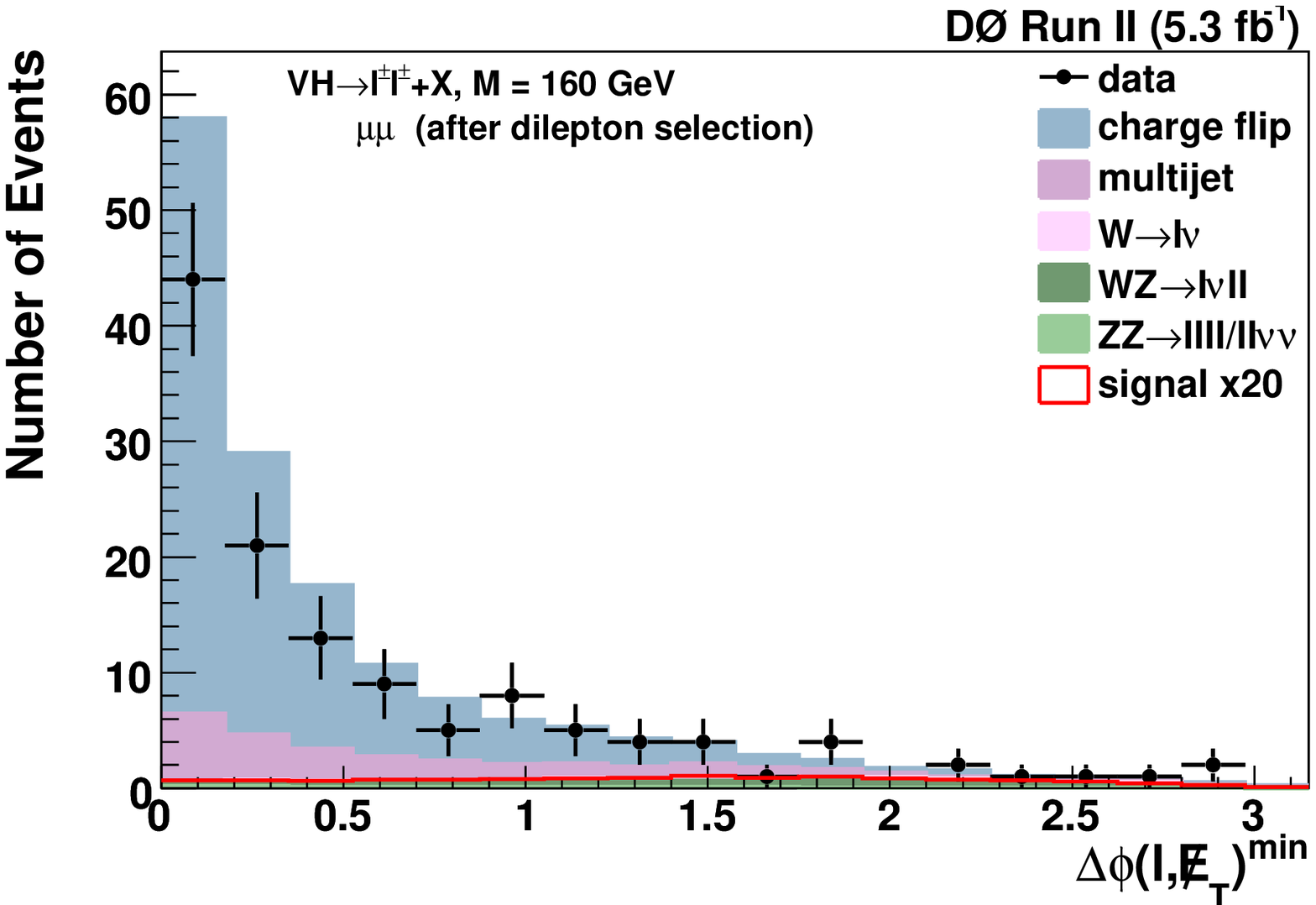}\\
\end{tabular}
\caption{(Color online) The distribution of (a) $M(\ell,\ell)$, (b) $\Delta R(\ell,\ell)$, (c) $N_{\rm jet}$,
         (d) $p_{T}^{\Sigma {\rm obj}}$, (e) $\met^{\mu}$, (f) $\Delta\phi(\ell,\met)^{\rm min}$, for the $\mu\mu$ channel
  comparing data and predicted backgrounds as well as the Higgs boson signal ($M_{H}$ = 160~GeV)
  expectation after the kinematic selection of like charge dilepton events. }
\label{fig:kine_mm}
\end{center}
\vspace{-2mm}
\end{figure*}

\begin{figure*}[!]
\begin{center}
\begin{tabular}{cc}
{\scriptsize\bf (a)}& {\scriptsize\bf (b)}\\
\vspace{-1mm}
\includegraphics[scale=0.39]{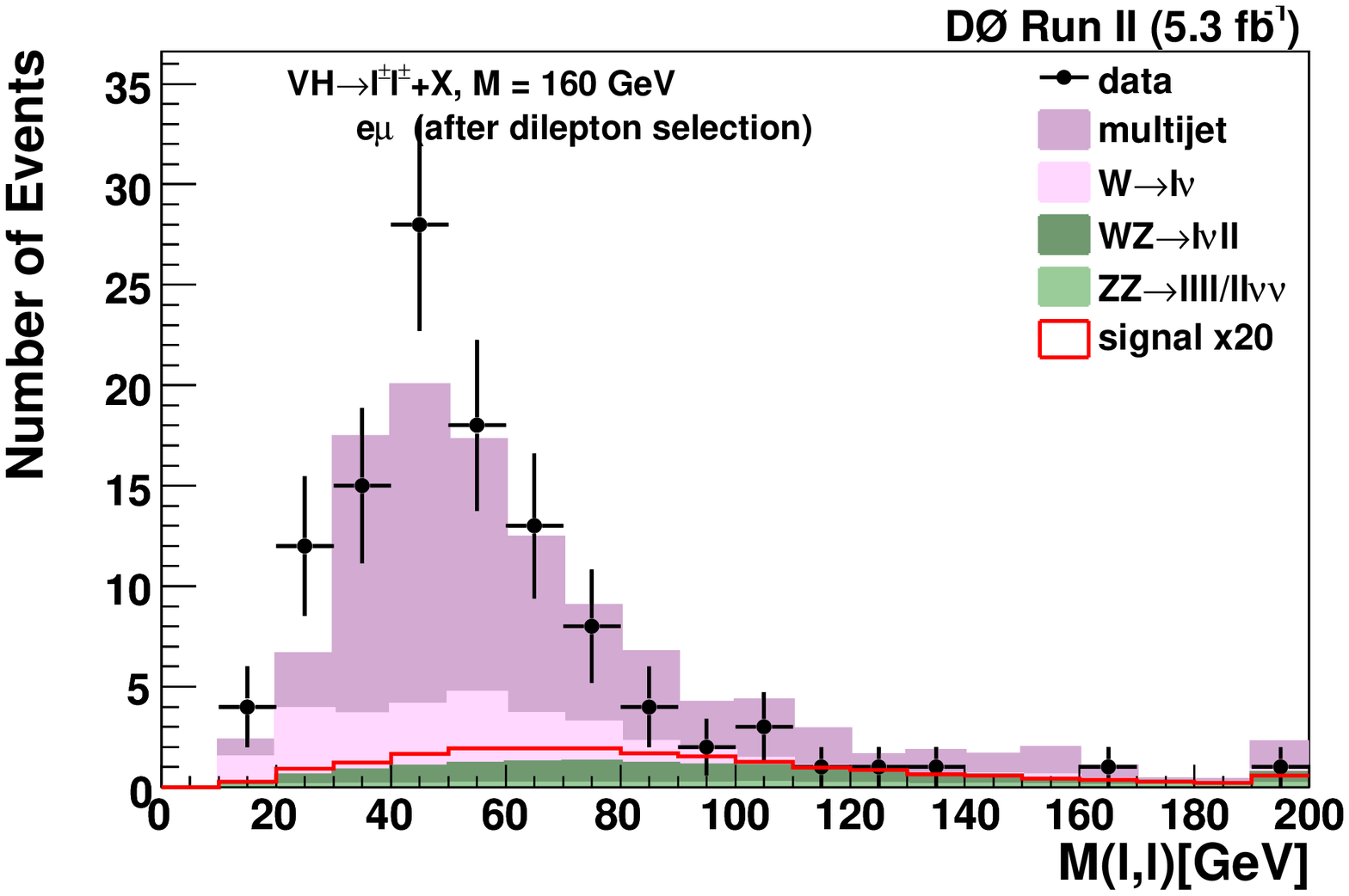}
&\includegraphics[scale=0.39]{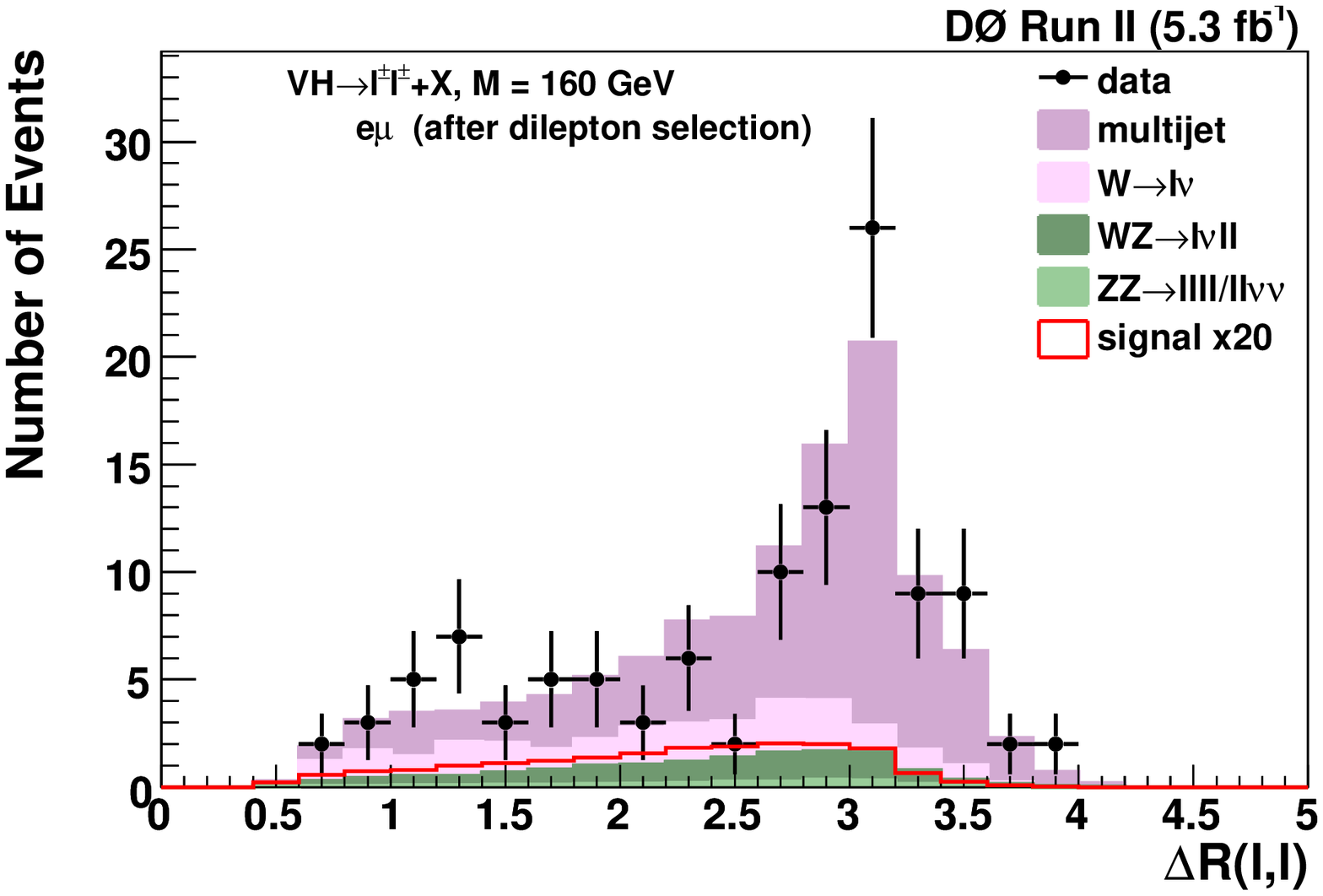}\\
{\scriptsize\bf (c)}& {\scriptsize\bf (d)}\\
\vspace{-1mm}
 \includegraphics[scale=0.39]{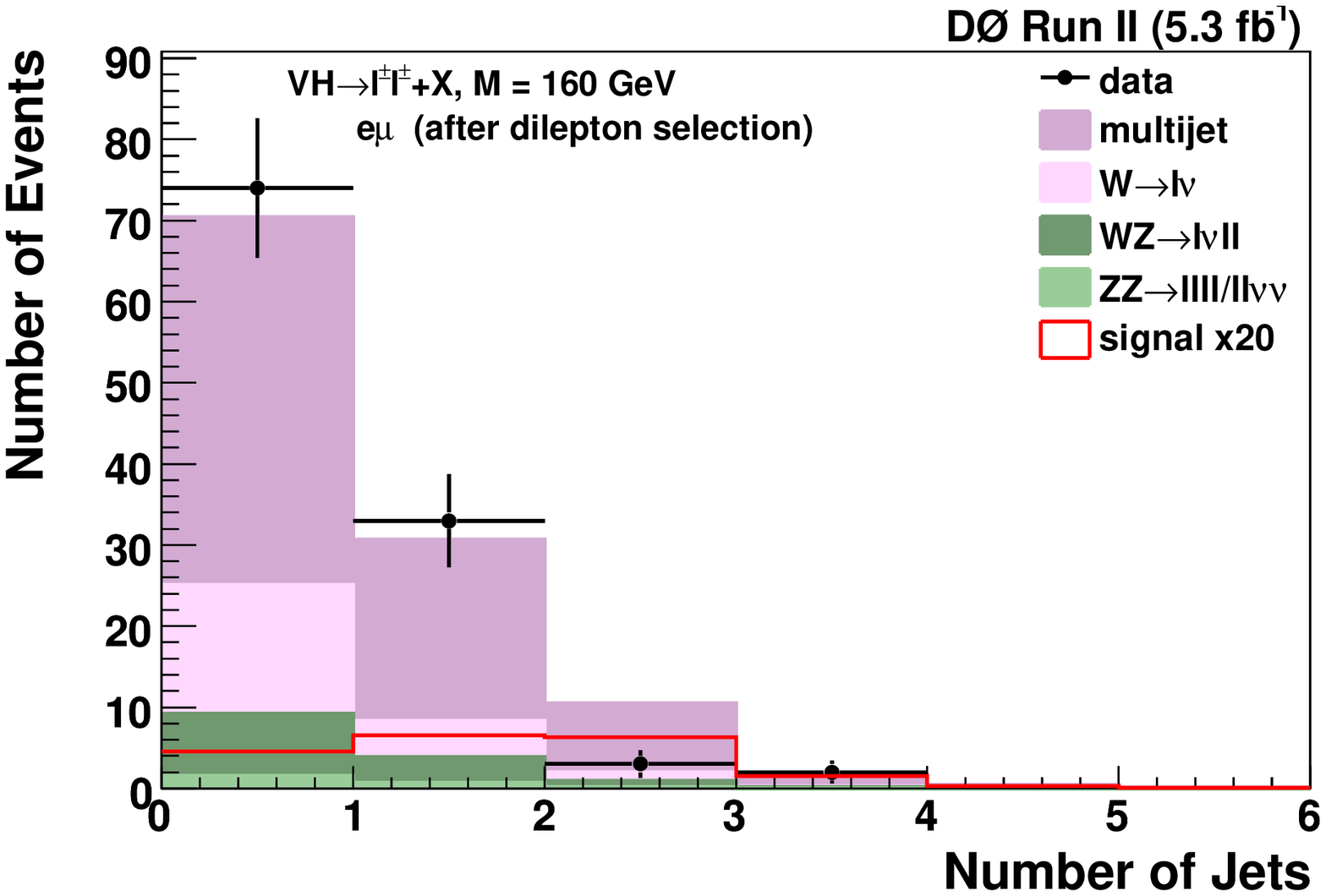}
&\includegraphics[scale=0.39]{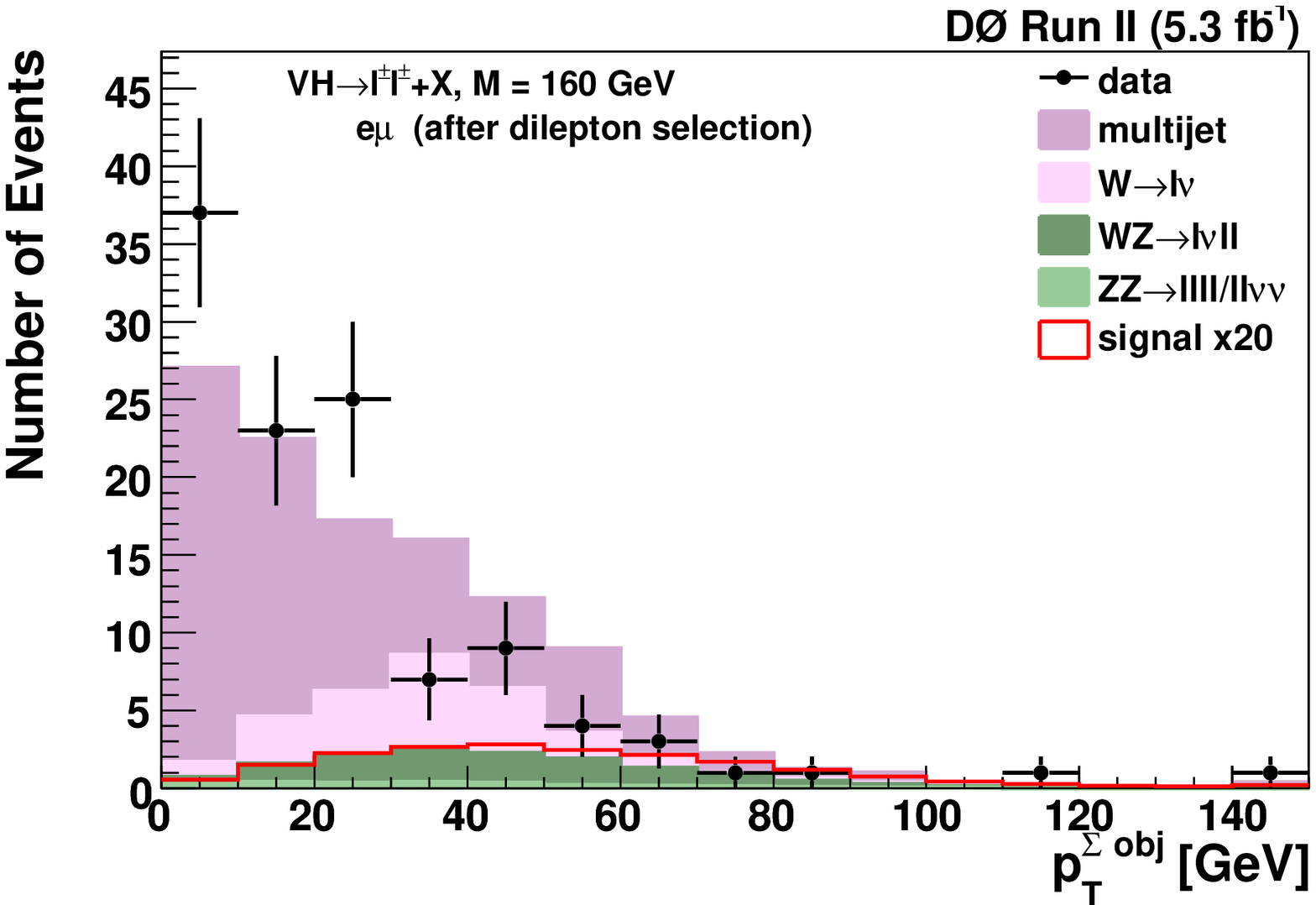}\\
{\scriptsize\bf (e)}& {\scriptsize\bf (f)}\\
\vspace{-1mm}
 \includegraphics[scale=0.39]{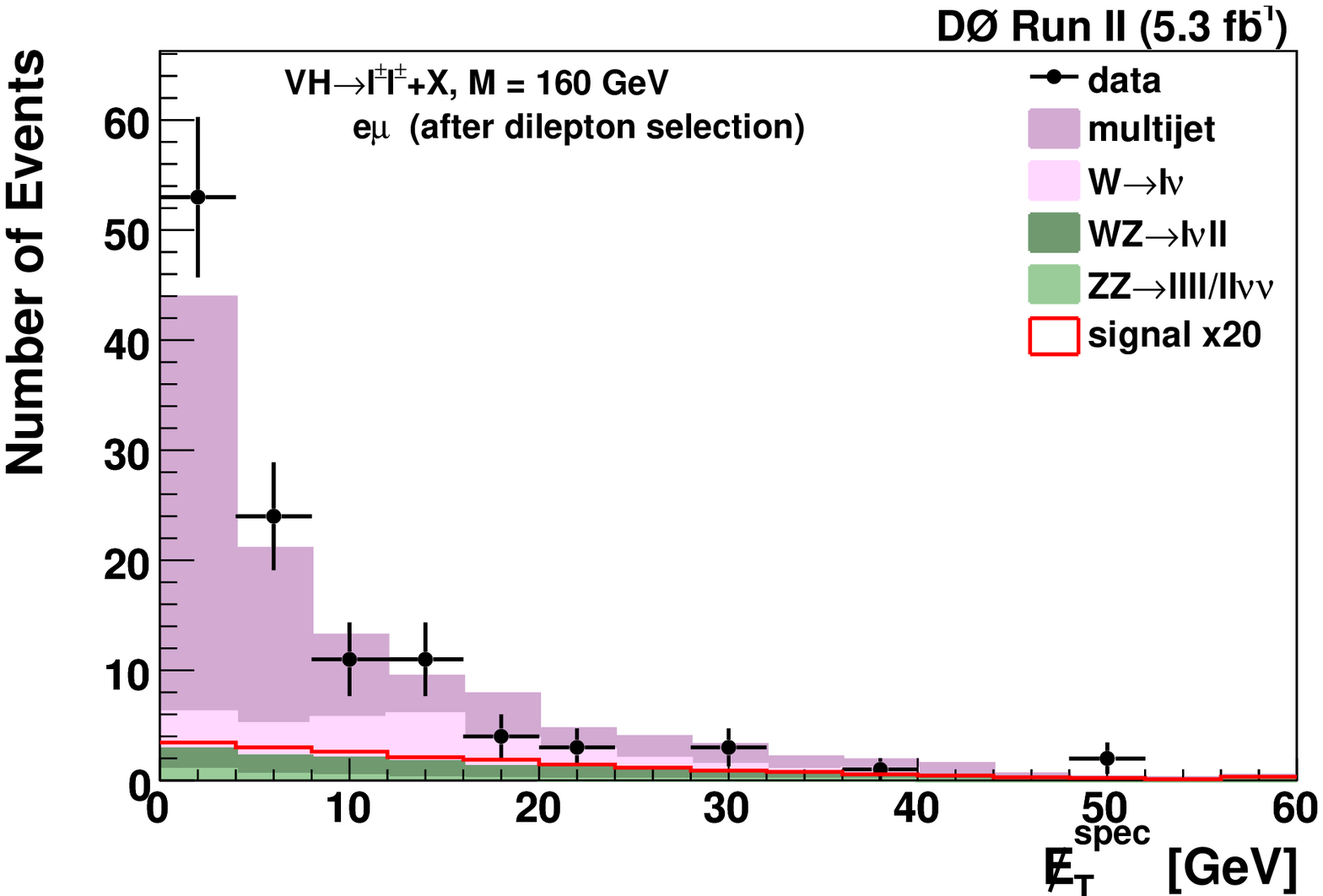}
&\includegraphics[scale=0.39]{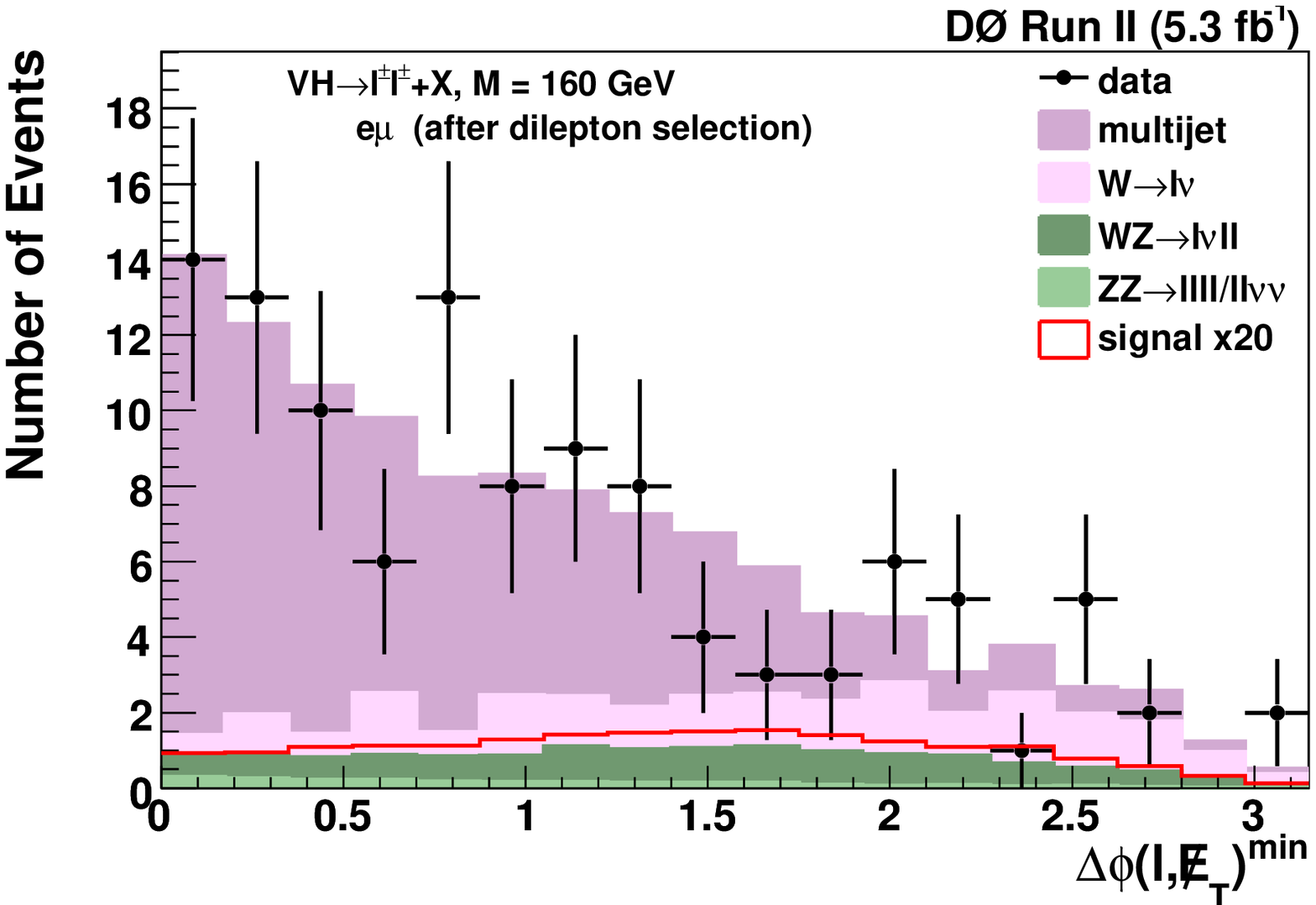}\\
\end{tabular}
\caption{(Color online) The distribution of (a) $M(\ell,\ell)$, (b) $\Delta R(\ell,\ell)$, (c) $N_{\rm jet}$,
         (d) $p_{T}^{\Sigma {\rm obj}}$, (e) $\met^{\rm spec}$, (f) $\Delta\phi(\ell,\met)^{\rm min}$, for the $e\mu$ channel
  comparing data and predicted backgrounds as well as the Higgs boson signal ($M_{H}$ = 160~GeV)
  expectation after the kinematic selection of like charge dilepton events.
  Due to the absence of charge flip background in $e\mu$ channel, the sample after
  kinematic selection is dominated by multijet and $W$+jet backgrounds.
  }
\label{fig:kine_em}
\end{center}
\vspace{-2mm}
\end{figure*}

The kinematic variables considered for the BDT inputs are:
\begin{itemize}
\item   Dilepton kinematics: \\
	leading and trailing lepton transverse momenta [{$p_{T}^{\ell1}$}, {$p_{T}^{\ell2}$}], 
	invariant mass [{$M(\ell,\ell)$}],
	angular separation [{$\Delta\eta(\ell,\ell)$}, {$\Delta\phi(\ell,\ell)$}, 
	{$\Delta R(\ell,\ell)$}];
\item   Kinematics of all leptons in the event:\\
          lepton multiplicity [{$N^{\ell}$}], vector sum and scalar sum of \PeTe~of all leptons
          [{$p_{T}^{\Sigma\ell}$}, {$\Sigma p_{T }^{\ell}$}];
\item   Kinematics of all jets in the event:\\
          jet multiplicity [{$N^{\rm jet}$}], vector sum and scalar sum of \PeTe~of all jets
          [{$H_{T}$}, {$\Sigma p_{T}^{\rm jet}$}];
\item   Kinematics of all the objects (leptons and jets) in the event:\\
          object multiplicity [{$N^{\rm obj}$}], vector and scalar sum of \PeTe~of all objects 
          [{$p_{T}^{\Sigma {\rm obj}}$}, {$\Sigma p_{T}^{\rm obj}$}];
\item   Missing transverse energy:\\
          missing transverse energy [{\met}], component perpendicular to the object/muon which is 
          closest to the \met~axis in $\phi$ [{\met$^{{\rm spec}/\mu}$}] 
          to be insensitive to a possible mismeasurement of the object momentum;
\item   Dilepton - \met~relation:\\
          transverse mass with minimum/maximum value
          [{$M_{T}(\ell,\met)^{\rm min/max}$}] 
          (those calculated with respect to electron/muon for $e\mu$ channel only
          [{$M_{T}(e/\mu,\met)$}]),
          minimum/maximum azimuthal angular separation [{$\Delta\phi(\ell,\met)^{\rm min/max}$}].
\end{itemize}
Between 11 and 17 variables are selected from this list for the training of BDTs
based on the discrimination power of the variables for a given channel.
Distributions of representative variables are shown in Figs.~\ref{fig:kine_ee}-\ref{fig:kine_em}
for $ee$, $\mu\mu$ and $e\mu$ channels after kinematic selection of like charge dilepton events.

\begin{figure*}[th]
\begin{center}
\begin{tabular}{cc}
{\scriptsize\bf (a)}& {\scriptsize\bf (b)} \\
\vspace{-1mm}
\includegraphics[scale=0.40]{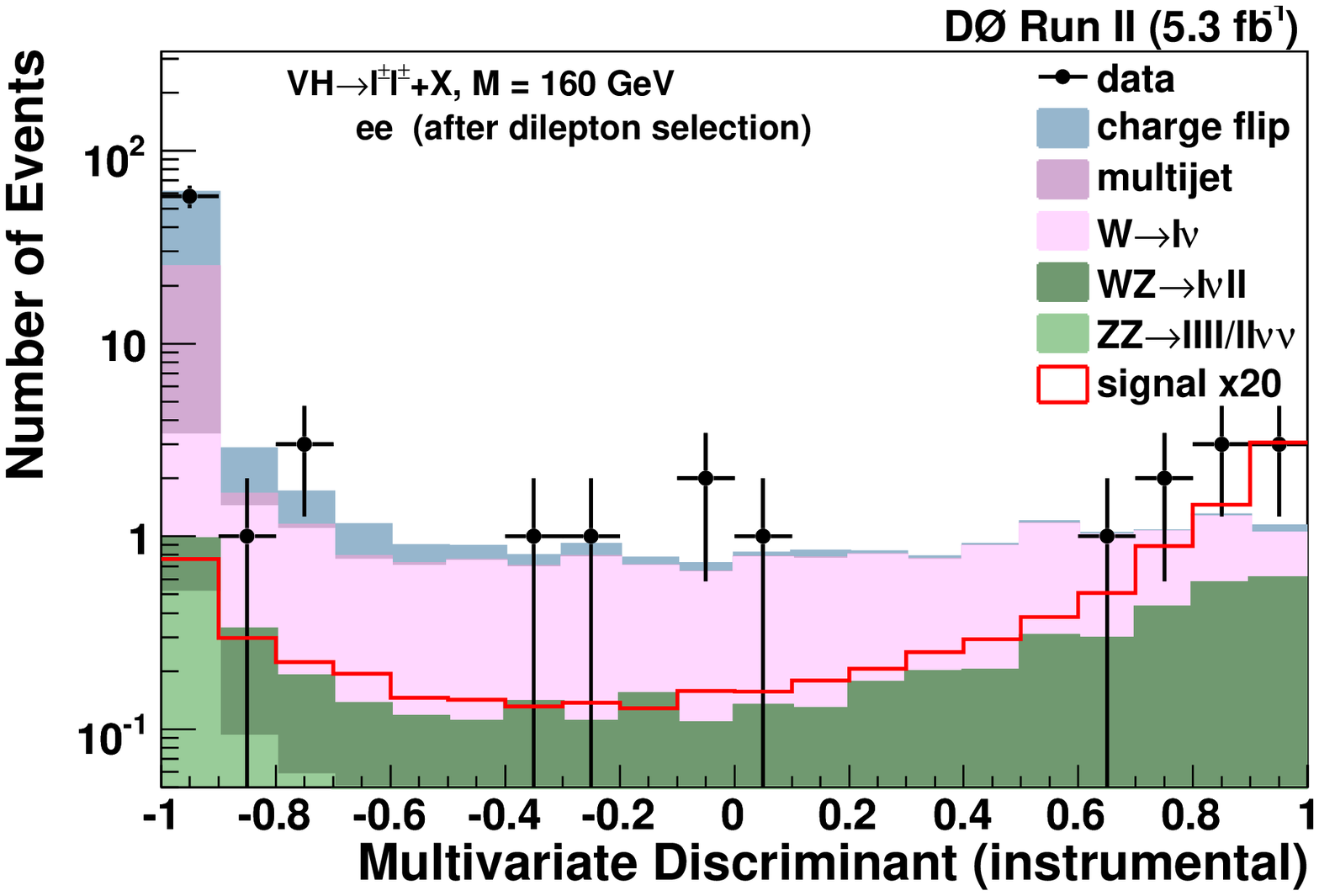}
&\includegraphics[scale=0.40]{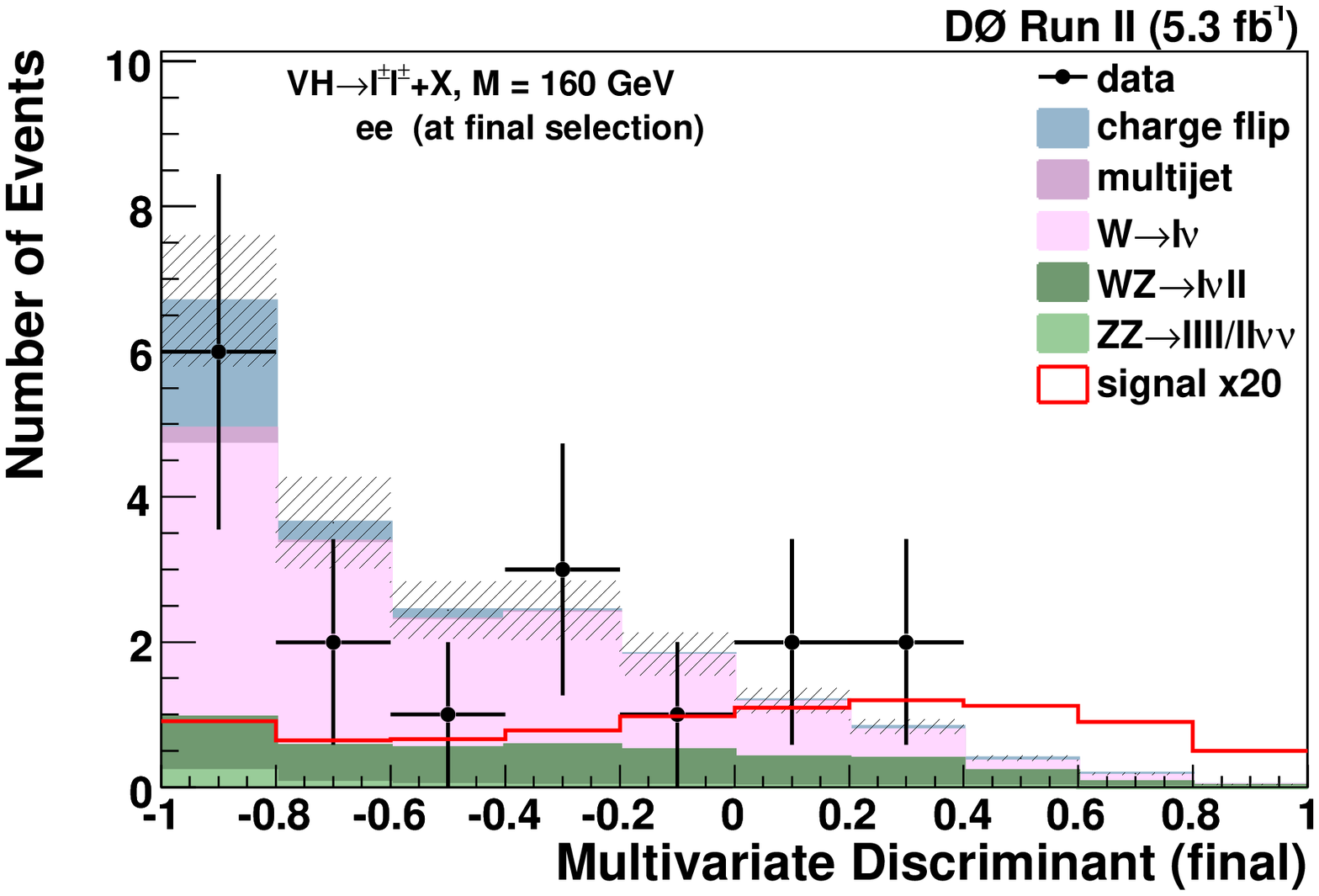}\\
{\scriptsize\bf (c)}& {\scriptsize\bf (d)}\\
\vspace{-1mm}
 \includegraphics[scale=0.40]{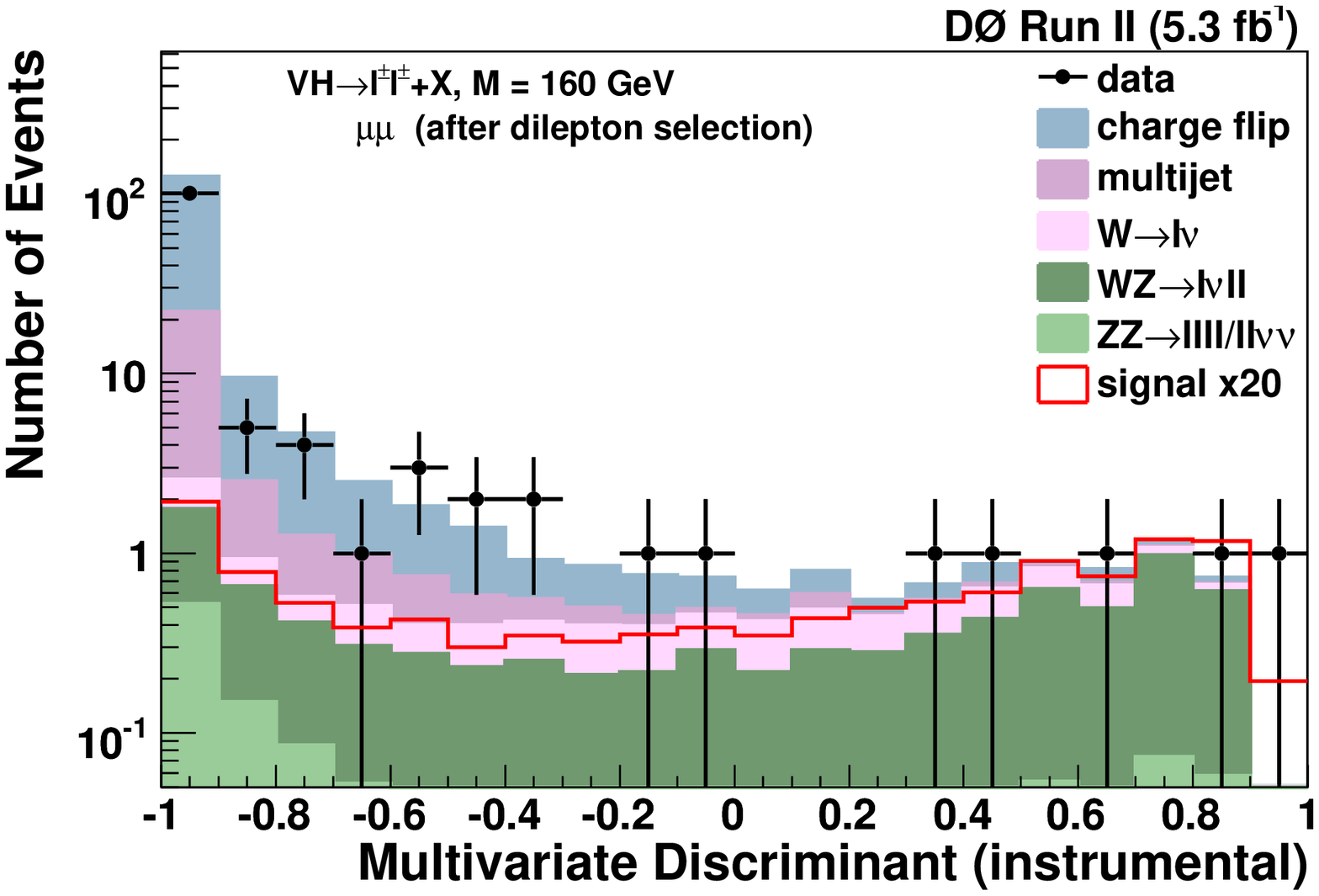}
&\includegraphics[scale=0.40]{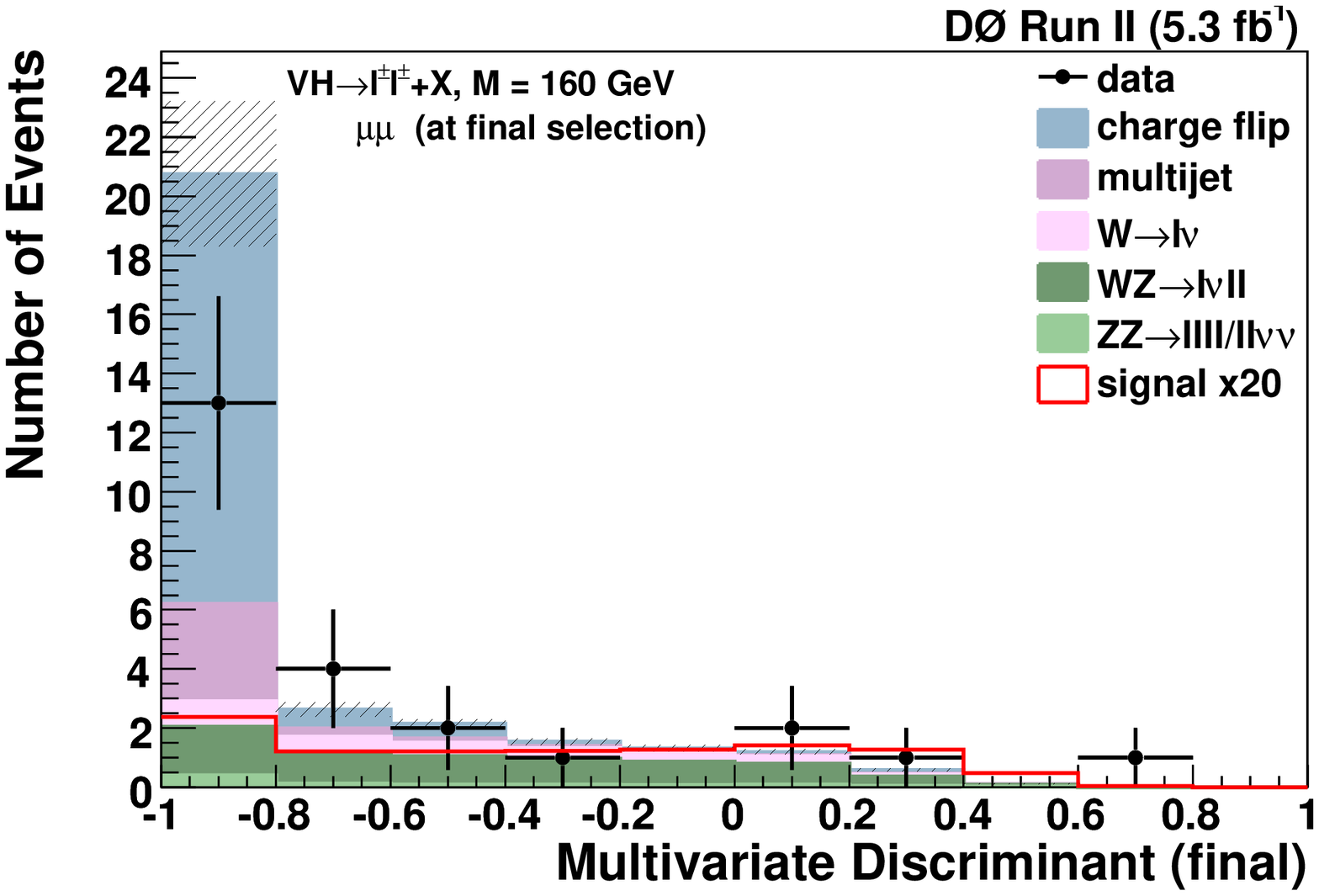}\\
{\scriptsize\bf (e)}& {\scriptsize\bf (f)}\\
\vspace{-1mm}
 \includegraphics[scale=0.40]{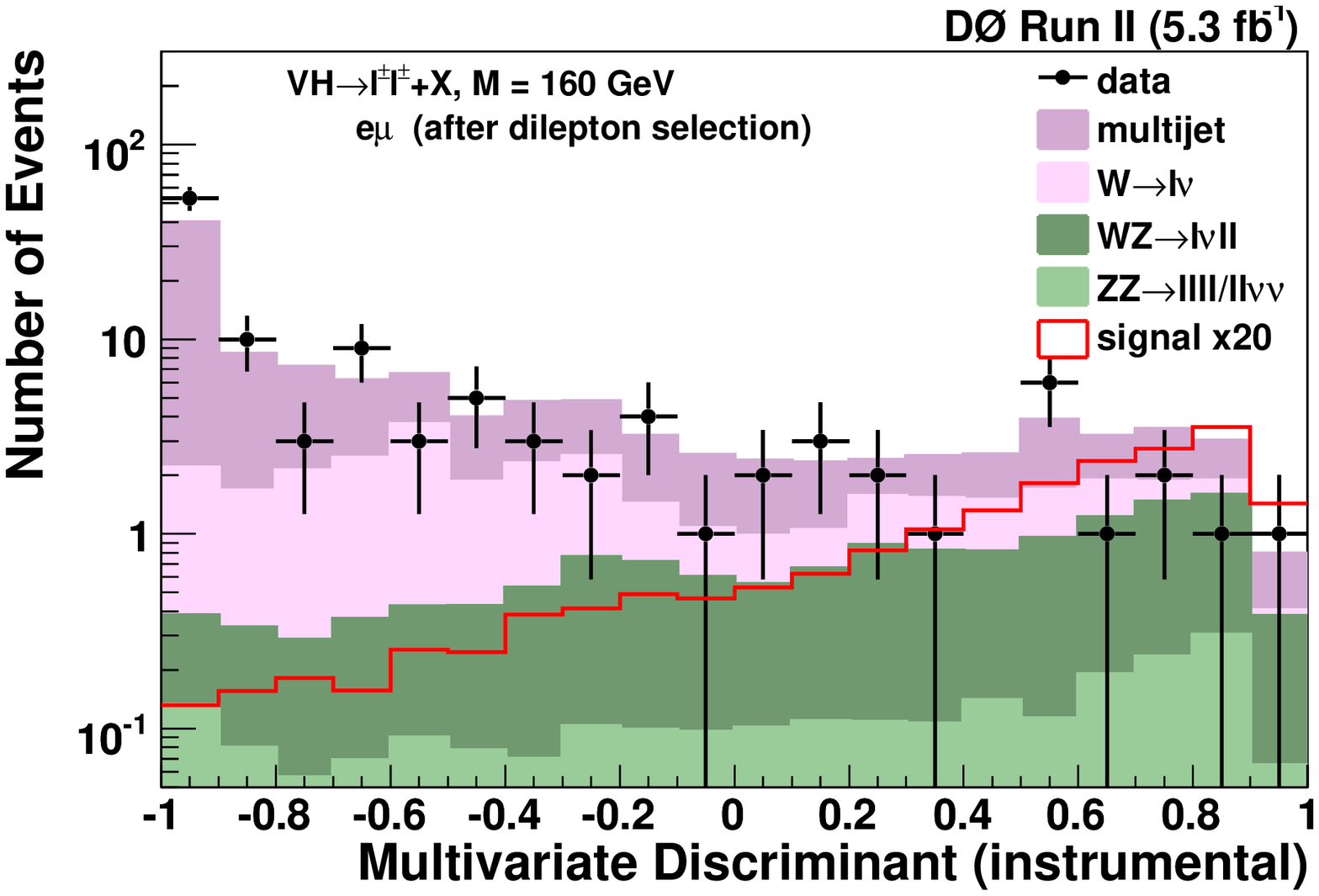}
&\includegraphics[scale=0.40]{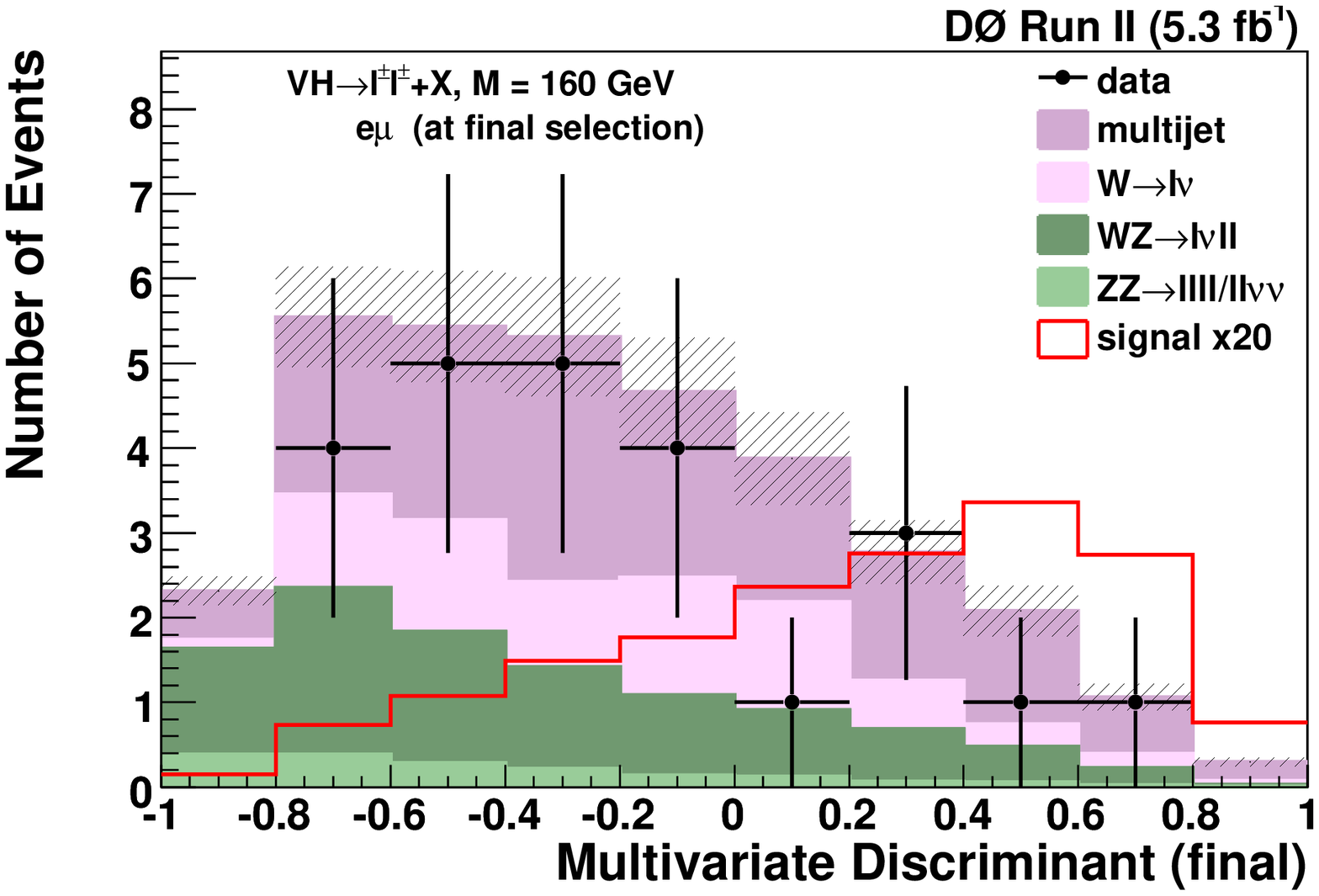}\\
\end{tabular}
\caption{(Color online) The distribution of (a, c, e) the multivariate discriminant against instrumental backgrounds, 
	    BDT$_i$, 
             and (b, d, f) the final discriminant which is an effective product of the 
             BDT outputs from the first stage (instrumental) and the second stage (physics)
	     for the (a, b) $ee$ (c, d) $\mu\mu$ and (e, f) $e\mu$ channels. 
             Data and background predictions corresponding to an integrated luminosity of 5.3~fb$^{-1}$ 
             and the signal distributions are shown for an assumed Higgs boson mass of 160~GeV.
             The shaded region represents the total uncertainty on the background prediction.
             The final discriminant distributions are shown after requirements on
             BDT$_i$ $>$ $-0.85$, $-0.9$, $-0.2$
              for the $ee$, $\mu\mu$, $e\mu$ channels, respectively,
             corresponding to approximately 90\% signal efficiency points for a Higgs boson mass of 160~GeV. 
	   }
\label{fig:bdt}
\end{center}
\end{figure*}

The BDT is trained for each Higgs boson mass considered.
The Higgs signal and background model samples described in the previous sections
are separated into two orthogonal samples such that there is no overlap between 
events used for training and those used to derive the final result of the search.
The training is carried out in two stages. 
The first stage uses one or two specific background processes that are dominant after the kinematic selection; 
charge flip and multijet for $ee$ and $\mu\mu$, multijet and $W$ + jet for the $e\mu$ channel.
The resulting instrumental BDT discriminant, BDT$_{i}$, is used to separate signal-like 
and background-like events (Fig.~\ref{fig:bdt} a, c and e).
The second stage of training, referred to as ``physics BDT'', uses those events that appear 
in the signal-like region of BDT$_{i}$,
where the threshold is optimised for a signal efficiency of approximately 90\% for each mass point. 
The combined background samples are used for $ee$ and $\mu\mu$ training, 
which are mostly diboson and $W$ boson or charge flip processes, and diboson only for $e\mu$.
All variables from the list that have discrimination power and are well modeled
by the prediction are selected for each stage of the training and for each of the three channels.
The input variables used for the physics BDT training for $ee$ and $\mu\mu$ channels are:
$p_{T}^{\ell1}$, $p_{T}^{\ell2}$, $\Delta\eta(\ell,\ell)$, $M(\ell,\ell)$,
$p_{T}^{\Sigma\ell}$, $\Sigma p_{T}^{\ell}$, 
\met, \met$^{\rm spec}$, $M_{T}(\ell,\met)^{\rm min}$, $\Delta\phi(\ell,\met)^{\rm min/max}$.
The input variables used for the physics BDT training for $e\mu$ channel are:
$p_{T}^{\ell1}$, $p_{T}^{\ell2}$, $\Delta\eta(\ell,\ell)$, $\Delta\phi(\ell,\ell)$, $\Delta R(\ell,\ell)$, 
$M(\ell,\ell)$, $p_{T}^{\Sigma\ell}$, 
$H_{T}$, $\Sigma p_{T}^{\rm jet}$, $p_{T}^{\Sigma {\rm obj}}$, $\Sigma p_{T}^{\rm obj}$,
\met, \met$^{\rm spec}$, $M_{T}(\ell,\met)^{\rm min/max}$, $\Delta\phi(\ell,\met)^{\rm min/max}$,
BDT$_{i}$.
The final discriminant for each channel is computed as an effective product of the two discriminants
after the selection based on BDT$_{i}$ (Fig.~\ref{fig:bdt}).

\section{\label{sec:result}CROSS SECTION LIMITS}

The final multivariate discriminants after all selection criteria 
(Fig.~\ref{fig:bdt} b, d and f) show that 
the data are well described by the sum of the background predictions. 
In absence of an excess in the number of observed events over the 
SM backgrounds, upper limits on the production cross 
section have been determined.

Uncertainties on the SM cross section for the processes
modeled by MC simulation are
5\% for the associated Higgs boson production, 
7\% for diboson ($WZ/ZZ$) production, and 6\% for $W$ boson production.
Experimental uncertainties assigned to the MC include
a normalization uncertainty of 4.7\%, which enfolds the lepton trigger and
identification efficiencies and their kinematic dependences and the uncertainty of 
the $Z$ boson production cross section used to study the normalization,
and a 2\% uncertainty on mismodeling of jets. 
An additional uncertainty is assigned to the $W$ boson background based on 
studies using the control samples; 20\% uncertainty is used for each misidentified muon,
50\%/14\% for each misidentified electron originating from a photon/jet.   
The instrumental backgrounds, charge flip and multijet, which are estimated directly from data,
have uncertainties between 11\%-42\%, arising mainly from the limited number
of events in the control samples and the extrapolation to the signal search region.
The mismodeling of the kinematics of the instrumental backgrounds are 
represented by a difference in the shape of the multivariate discriminants,
and corresponds to up to 20\% uncertainty on the final yield. 
The uncertainties described above are considered uncorrelated.

The number of predicted and observed events after the kinematic selection of two
like charged leptons and after the additional selection based on the multivariate discriminant BDT$_i$
are listed in Table~\ref{tab:events}.
The total uncertainties associated with each background and signal processes, 
excluding the shape uncertainty for the instrumental backgrounds, are given.

\begin{table*}
\caption{\label{tab:events} 
	The number of predicted and observed events for 5.3~fb$^{-1}$ of Run II integrated luminosity,
	after kinematic selection of like charged dileptons 
	and after the final selection based on the multivariate discriminant against instrumental background, BDT$_i$.
	This excludes the two control regions used for the estimation of the charge flip and multijet backgrounds.
	The uncertainties reflect the total systematic uncertainties.}
\begin{ruledtabular}
\begin{tabular}{l c | c r c l c r c l c r c l c | c r c l c r c l c r c l }
  && \multicolumn{13}{c|}{kinematic selection}    & \multicolumn{12}{c}{kinematic $+$ BDT$_{i}$ selection }  \\
  &&& \multicolumn{3}{c}{$ee$} && \multicolumn{3}{c}{$\mu\mu$} && \multicolumn{3}{c}{$e\mu$} 	
  &&& \multicolumn{3}{c}{$ee$} && \multicolumn{3}{c}{$\mu\mu$} && \multicolumn{3}{c}{$e\mu$} \\

\hline\hline
$WZ\rightarrow\ell\nu\ell\ell$ 	&&& 4.46 & $\pm$ & 0.39 && 7.56 & $\pm$ & 0.66 && 11.81 & $\pm$ & 1.03 
						&&& 3.89 & $\pm$ & 0.34 && 6.32 & $\pm$ & 0.55 && 8.98 & $\pm$ & 0.78\\
$ZZ\rightarrow\ell\ell\ell\ell$ 	&&& 0.92 & $\pm$ & 0.08 && 1.38 & $\pm$ & 0.12 && 2.34 & $\pm$ & 0.20 
						&&& 0.35 & $\pm$ & 0.03 && 0.86 & $\pm$ & 0.08 && 1.66 & $\pm$ & 0.14\\
$W\rightarrow\ell\nu$ 		&&& 14.8 & $\pm$ & 3.3 && 3.9 & $\pm$ & 0.8 && 21.4 & $\pm$ & 6.7 
						&&& 12.8 & $\pm$ & 2.9 && 3.0 & $\pm$ & 0.6 && 7.2 & $\pm$ & 2.3\\
multijet 					&&& 22.0 & $\pm$ & 7.5 && 23.5 & $\pm$ & 7.6 && 78.6 & $\pm$ & 16.7 
						&&& 0.3 & $\pm$ & 0.1 && 3.9 & $\pm$ & 1.3 && 15.5 & $\pm$ & 3.3\\
charge flip 				&&& 39.0 & $\pm$ & 6.9 && 118.4 & $\pm$ & 17.0 && $-$ & $\pm$ & $-$ 
						&&& 2.4 & $\pm$ & 0.4 && 16.2 & $\pm$ & 2.5 && $-$ & $\pm$ & $-$\\
\hline
total background 			&&& 81.2 & $\pm$ & 10.7 && 154.7 & $\pm$ & 18.7 && 114.2 & $\pm$ & 18.1 
						&&& 19.7 & $\pm$ & 2.9 && 30.3 & $\pm$ & 2.9 && 33.4 & $\pm$ & 4.1\\
\hline\hline
data 						&&& \multicolumn{3}{c}{76} && \multicolumn{3}{c}{125} && \multicolumn{3}{c}{112} 
						&&& \multicolumn{3}{c}{17} && \multicolumn{3}{c}{24} && \multicolumn{3}{c}{24}\\
\hline\hline
$VH$ ($M_{H}$ = 120 GeV) 	&&& 0.29 & $\pm$ & 0.02 && 0.44 & $\pm$ & 0.03 && 0.73 & $\pm$ & 0.05 
						&&& 0.26 & $\pm$ & 0.02 && 0.35 & $\pm$ & 0.03 && 0.65 & $\pm$ & 0.05\\
$VH$ ($M_{H}$ = 160 GeV) 	&&& 0.49 & $\pm$ & 0.03 && 0.62 & $\pm$ & 0.04 && 0.96 & $\pm$ & 0.07 
						&&& 0.44 & $\pm$ & 0.03 && 0.52 & $\pm$ & 0.04 && 0.86 & $\pm$ & 0.06\\
$VH$ ($M_{H}$ = 200 GeV) 	&&& 0.21 & $\pm$ & 0.01 && 0.27 & $\pm$ & 0.02 && 0.47 & $\pm$ & 0.03 
						&&& 0.19 & $\pm$ & 0.01 && 0.24 & $\pm$ & 0.02 && 0.42 & $\pm$ & 0.03\\
\end{tabular}
\end{ruledtabular}
\end{table*}

\begin{figure*}[htb!]
\begin{center}
\begin{tabular}{cc}
{\scriptsize\bf (a)}& {\scriptsize\bf (b)}\\
\includegraphics[width=.51\textwidth]{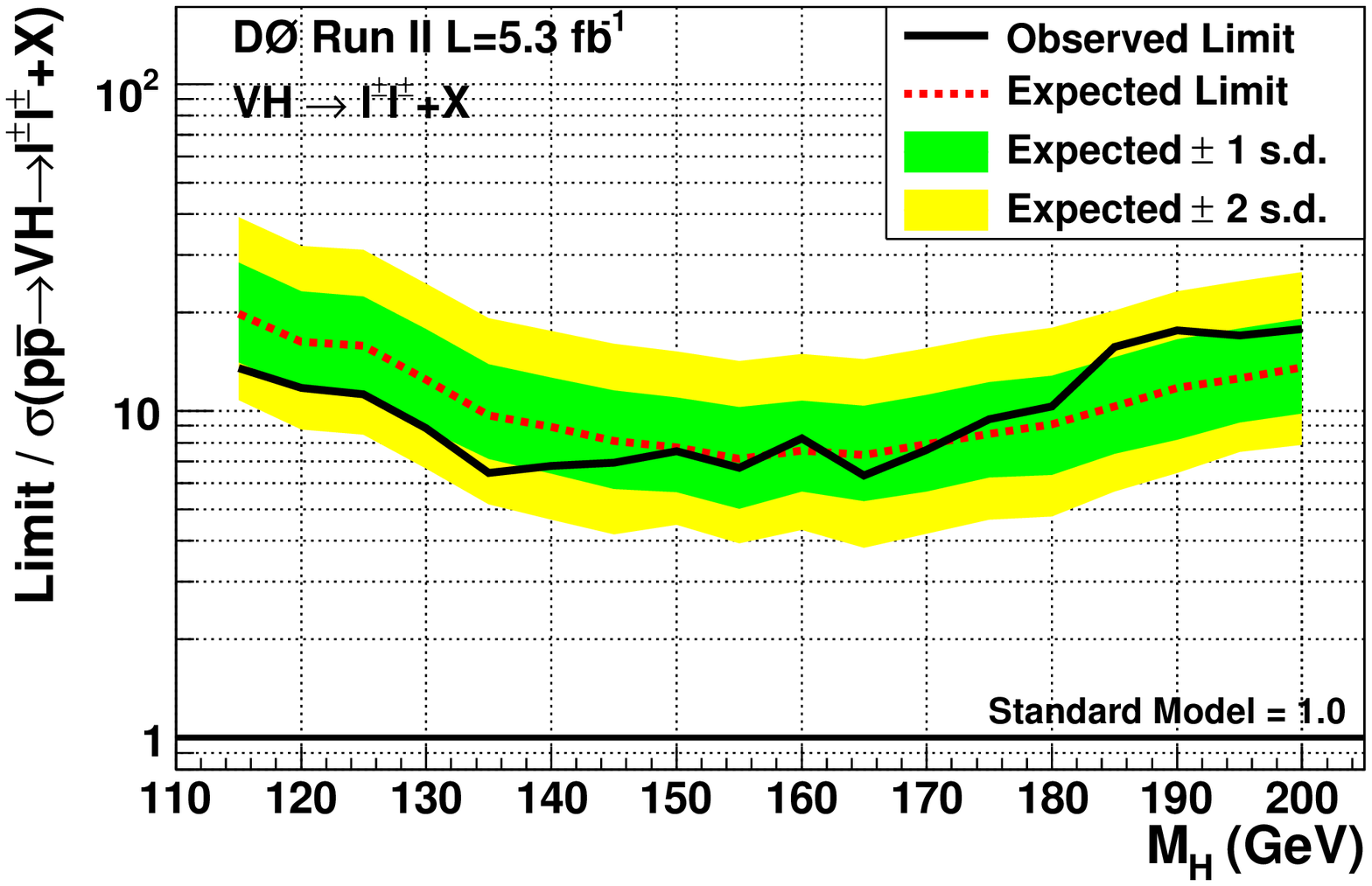}
&\includegraphics[width=.48\textwidth]{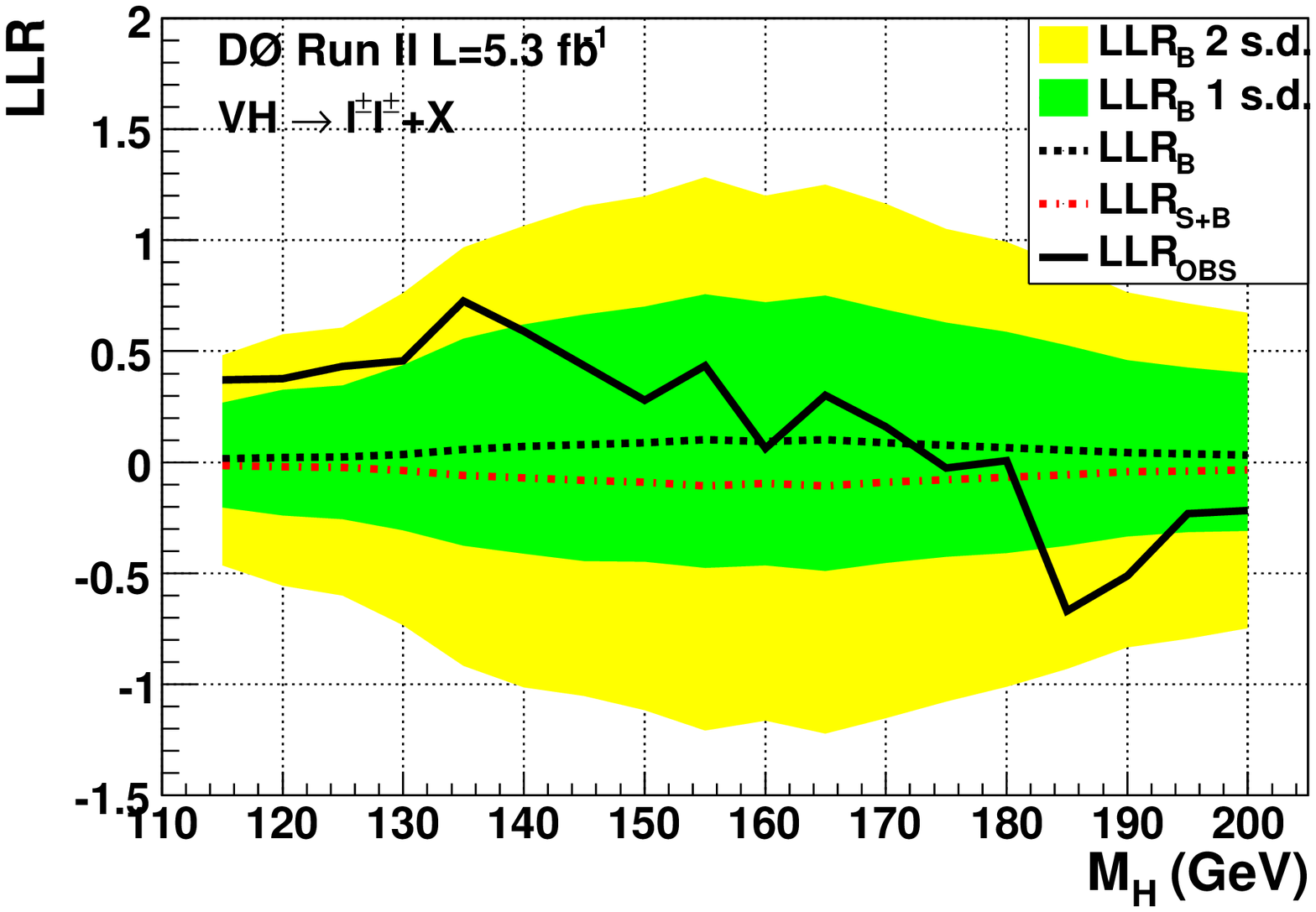}\\
\end{tabular}
\caption{(Color online) (a) the expected and observed production cross section limits in terms of the 
	ratio to the SM cross section as a function of the Higgs boson mass
	for an integrated luminosity of 5.3~fb$^{-1}$ combining 
	$ee$, $\mu\mu$ and $e\mu$ channels.
	(b) the corresponding LLR distribution.
	}
\label{fig:limits_sm}
\end{center}
\vspace{0.2em}
\end{figure*}

Cross section upper limits have been determined with
the modified frequentist approach~\cite{bib:limit}
with a log-likelihood ratio (LLR) test statistics. 
The systematic uncertainty of the signal and background predictions are represented
by a Gaussian distributed fluctuation of the expected yield, where correlations
across different channels for any particular uncertainty are taken into account. 
The details of the calculation are explained in Ref.~\cite{bib:hww}.
Upper limits on the $W/ZH$ cross section expressed as a ratio to the SM 
Higgs cross section and the corresponding LLR distributions are shown in Fig.~\ref{fig:limits_sm}.
Ratio limits for each dilepton channel and for the combination are summarized in Table~\ref{tab:limits_sm}.

\begin{table*}[tbh]
\caption{\label{tab:limits_sm} 
	The expected and observed production cross section limits in terms of 
	ratios to the SM cross section for individual channels and for the combination.
	}
\begin{ruledtabular}
{\fontsize{9}{11}\selectfont
\begin{tabular}{c|c|cccccccccccccccccc}
\multicolumn{20}{c}{D0  Run II Limits for $W/ZH \rightarrow \rightarrow \ell^\pm\ell'^\pm + X$ (5.3~fb$^{-1}$)}\\
\hline
\multicolumn{2}{c|}{$m_{H}$} 	&115 &120 &125 &130 &135 &140 &145 &150 
						&155 &160 &165 &170 &175 &180 &185 &190 &195 &200 \\
\hline\hline
$ee$     &exp &44.6   &38.4   &30.6   &25.4   &22.3   &19.1   &18.2   &17.4   &15.9   &16.3   &16.9   &17.8   &18.8   &20.1   &22.6   &25.6   &27.7   &29.7\\
         &obs &27.0   &29.8   &21.9   &20.3   &18.2   &17.9   &19.5   &22.6   &16.9   &17.1   &18.8   &20.8   &26.8   &26.6   &34.9   &45.3   &39.7   &40.4\\
$\mu\mu$ &exp &37.6   &30.0   &33.5   &22.8   &20.2   &18.2   &17.8   &17.3   &16.7   &17.0   &17.6   &20.1   &20.9   &23.0   &26.0   &30.0   &33.2   &34.5\\
         &obs &34.0   &25.3   &30.4   &23.9   &19.6   &18.2   &19.1   &18.7   &20.4   &22.0   &19.4   &23.3   &25.4   &24.8   &35.1   &40.2   &39.7   &37.6\\
$e\mu$   &exp &27.7   &23.0   &23.3   &20.3   &13.9   &13.4   &11.6   &10.9   &10.3   &11.5   &10.5   &11.3   &12.7   &13.2   &15.0   &17.2   &18.5   &19.6\\
         &obs &21.9   &17.2   &17.8   &12.6   &8.7    &9.0    &8.1    &8.0    &7.3    &9.6    &7.1    &8.0    &9.1    &11.7   &14.8   &15.6   &19.3   &22.0\\
\hline
all      &exp &19.8   &16.2   &15.8   &12.5   &9.7    &8.9    &8.1    &7.7    &7.1    &7.6    &7.3    &7.9    &8.5    &9.1    &10.3   &11.7   &12.6   &13.6\\
         &obs &13.5   &11.8   &11.2   &8.8    &6.4    &6.8    &7.0    &7.5    &6.7    &8.2    &6.4    &7.6    &9.4    &10.3   &15.7   &17.7   &17.0   &17.8\\
\end{tabular}
}
\end{ruledtabular}
\end{table*}

\section{\label{sec:conclusion}CONCLUSIONS}

A search for associated production of SM Higgs boson,
 $p\bar{p} \rightarrow W/ZH$, has been performed with a final state
with two like charged leptons, $W/ZH \rightarrow \ell^{\pm}\ell'^{\pm} + X$, in the
$ee$, $e\mu$ and $\mu\mu$ channels. After all selection criteria, 17
events in the $ee$ channel, 24 events in the $e\mu$ channel, and 24
events in the $\mu\mu$ channel have been observed in agreement with
SM predictions. The observed (expected) upper
limits on $\sigma(W/ZH)~\times~{\cal B}(W/ZH\rightarrow \ell^\pm\ell'^\pm + X)$
for all three channels combined
using a total integrated luminosity of 5.3~fb$^{-1}$ 
collected by the D0 detector in $p\bar{p}$ collisions at $\sqrt{s} = 1.96$ TeV,
expressed as a ratio to the SM Higgs cross section, 
are found to be 
6.4 (7.3) for a Higgs boson mass of 165~GeV and 13.5 (19.8) for 115~GeV.

\section{\label{sec:ackn}ACKNOWLEDGEMENT}
%
We thank the staffs at Fermilab and collaborating institutions,
and acknowledge support from the
DOE and NSF (USA);
CEA and CNRS/IN2P3 (France);
FASI, Rosatom and RFBR (Russia);
CNPq, FAPERJ, FAPESP and FUNDUNESP (Brazil);
DAE and DST (India);
Colciencias (Colombia);
CONACyT (Mexico);
KRF and KOSEF (Korea);
CONICET and UBACyT (Argentina);
FOM (The Netherlands);
STFC and the Royal Society (United Kingdom);
MSMT and GACR (Czech Republic);
CRC Program and NSERC (Canada);
BMBF and DFG (Germany);
SFI (Ireland);
The Swedish Research Council (Sweden);
and
CAS and CNSF (China).



\begin{thebibliography}{99}

\bibitem{prev-version-pub} 
 V.~M.~Abazov {\sl et al.} [D0 Collaboration],
Phys.\ Rev.\ Lett.\ {\bf97}, 151804 (2006).

\bibitem{cdf-hww}
T. Aaltonen {\sl et al.} [CDF Collaboration], 
Phys.\ Rev.\ Lett.\ {\bf 104}, 061803 (2010).

\bibitem{tevcomb-hww}
T. Aaltonen {\sl et al.} [CDF and D0 Collaborations],
Phys.\ Rev.\ Lett.\ {\bf 104}, 061802 (2010).

\bibitem{bib:d0det}
  V.~M.~Abazov {\sl et al.}  [D0 Collaboration],
  Nucl.\ Instrum.\ Meth.\  A {\bf 565}, 463 (2006).

\bibitem{bib:layer0}
  R.~Angstadt {\sl et al.}  [D0 Collaboration],
  Nucl.\ Instrum.\ Meth.\  A {\bf 622}, 298 (2010).
  
\bibitem{bib:eta}
The pseudorapidity is defined as $\eta = - \ln[\tan(\theta/2)]$,
where $\theta$ is the polar angle relative to the proton beam direction.

\bibitem{bib:jet}
G. C. Blazey {\sl et al.}, arXiv:hep-ex/0005012.

\bibitem{bib:geant}
R. Brun and F. Carminati, CERN Program Library Long
Writeup W5013, 1993 (unpublished).

\bibitem{bib:pythia}
  T.~Sj\"ostrand {\sl et al.}, Comp.\ Phys.\ Comm. {\bf 135}, 238 (2001);
  we use version 6.323 or later.

\bibitem{bib:cteq}
J.~Pumplin {\sl et al.}, J.\ High Energy Phys.\ {\bf 07}, 12 (2002).
  
\bibitem{bib:vh-xs}
J. Baglio and A. Djouadi, J.\ High Energy Phys.\ {\bf 10}, 064 (2010).
 
\bibitem{bib:hdecay}
  A.~Djouadi {\sl et al.}, Comp.\ Phys.\ Comm. {\bf 108}, 56 (1998);
  we use version 3.53.

\bibitem{bib:dibo-xs} 
  J.M. Campbell and R.K. Ellis, Phys. Rev. D {\bf 60}, 113006 (1999); 
  we use $\sigma(WW)$ = 11.66 pb, $\sigma(WZ)$ = 3.45 pb, and $\sigma(ZZ)$ = 1.37 pb.
  
\bibitem{bib:mcatnlo}
S. Frixione and B.R. Webber, J.\ High Energy Phys.\ {\bf 06}, 029 (2002).

\bibitem{bib:alpgen} 
M.L. Mangano {\sl et al.}, J.\ High Energy Phys.\ {\bf 07}, 001 (2003); we use version 2.11.

\bibitem{bib:v-xs} R.~Hamberg, W.~L.~van Neerven, and T.~Matsuura,
  Nucl.\ Phys. {\bf B359}, 343 (1991) [Erratum-ibid. {\bf B644}, 403 (2002)].
  
\bibitem{bib:tt-xs}
S. Moch and P. Uwer, Phys. Rev. D {\bf 78}, 034003 (2008);
we use $\sigma(tt)$ = 7.88 pb.

\bibitem{bib:bdt} 
L. Breiman {\sl et al.}, {\it Classification and Regression Trees} (Wadsworth, Belmont, CA, 1984); 
Proceedings of the Thirteenth International Conference, Bari, Italy, 1996, edited by L. Saitta (Morgan Kaufmann, San Francisco, 1996), p148.

\bibitem{bib:limit} T.~Junk, Nucl.\ Instrum.\ Meth.\ A {\bf 434}, 435 (1999).

\bibitem{bib:hww}
T. Aaltonen {\sl et al.} [CDF and D0 Collaborations], Phys. Rev. Lett. {\bf 104}, 061802 (2010).

\end{thebibliography}
\end{document}